\documentclass[11pt]{article}
\usepackage[top=1in, bottom=1in, left=1in, right=1in]{geometry}

\usepackage[utf8]{inputenc} 
\usepackage[T1]{fontenc}    
\usepackage{hyperref}       
\usepackage{url}            
\usepackage{booktabs}       
\usepackage{amsfonts}       
\usepackage{nicefrac}       
\usepackage{microtype}      
\usepackage{graphicx}
\usepackage{subfig}      
\usepackage{algorithm}
\usepackage{algpseudocode}
\usepackage{multirow}
\usepackage{array, makecell}
\usepackage{amssymb}
\usepackage{amsmath}
\usepackage{amsthm}
\usepackage{arydshln}
\usepackage{cleveref}
\crefformat{footnote}{#2\footnotemark[#1]#3}
\usepackage[numbers]{natbib}

\newcommand{\set}[1]{\left\{#1\right\}}

\renewcommand{\sim}{\textrm{sim}}
\newcommand{\DS}{\textrm{DS}}

\newcommand{\R}{\mathbb{R}}

\newtheorem{theorem}{Theorem}[section]
\newtheorem{remark}[theorem]{Remark}

\usepackage{tikz}
\newcommand{\greencircle}[2][black!30!green,fill=black!30!green]{\tikz[baseline=-0.5ex]\draw[#1,radius=#2] (0,0) circle ;}%
\newcommand{\whitecircle}[2][white,fill=white]{\tikz[baseline=-0.5ex]\draw[#1,radius=#2] (0,0) circle ;}%

\title{Dialect Diversity in Text Summarization on Twitter\footnote{Accepted to The Web Conference 2021.}}

\usepackage{authblk}

\author{Vijay Keswani}
\author{L. Elisa Celis}
\affil{Yale University}
\date{}

\begin{document}

\maketitle

\begin{abstract}
Discussions on Twitter involve participation from different communities with different dialects
and it is often necessary to summarize a large number of posts into a representative sample to provide a synopsis.
Yet, any such representative sample should sufficiently portray the underlying dialect diversity to present the voices of different participating communities representing the dialects. 
Extractive summarization algorithms perform the task of constructing subsets that succinctly capture the topic of any given set of posts.
However, we observe that there is dialect bias in the summaries generated by common summarization approaches, 
i.e., they often return summaries that under-represent certain dialects.

The vast majority of existing ``fair'' summarization approaches
require socially salient attribute labels (in this case, dialect) to ensure that the generated summary is fair with respect to the socially salient attribute.
Nevertheless, in many applications, these labels do not exist.
Furthermore, due to the ever-evolving nature of dialects in social media, it is unreasonable
to label or accurately infer the dialect of every social media post.
To correct for the dialect bias, we employ a framework that takes an existing text summarization algorithm as a blackbox and, using a small set of dialect-diverse sentences, returns a summary that is relatively more dialect-diverse.
Crucially, this approach does not need the posts {being summarized} to have dialect labels, ensuring that the diversification process is independent of dialect classification/identification models.
We show the efficacy of our approach on Twitter datasets containing posts written in dialects used by different social groups defined by race or gender; in all cases, our approach leads to improved dialect diversity compared to standard text summarization approaches.
\end{abstract}

\section{Introduction} \label{sec:introduction}
    
The popularity of social media has led to a centralized discussion on a variety of topics.
This has encouraged the participation of people from different communities in online discussions, helping induce a more diverse and robust dialogue, and giving voice to marginalized communities \cite{lavan2015negro}.
Twitter, for example, receives around 500 million posts per day, with posts written in more than 50 languages\footnote{\url{https://www.internetlivestats.com/twitter-statistics/}}. 
Within English, Twitter sees a large number of posts from different dialects; this diversity has even encouraged linguists to use Twitter posts to study dialects, for example, to map regional dialect variation \cite{huang2016understanding, doyle2014mapping} or to construct parsing tools for minority dialects \cite{blodgett2018twitter, jorgensen2015challenges}.
Yet, automated language tools are often unable to handle the dialect diversity in Twitter, leading to issues like disparate accuracy of language identification between posts written in African-American English (AAE) and standard English \cite{blodgett2017racial}, or dialect-based discrepancies in abusive speech detection \cite{sap2019risk, park2018reducing}.

Summarization algorithms for social media platforms, like Twitter, perform the task of condensing a large number of posts into a small representative sample.
They are useful because they provide the users with a synopsis of long discussions on these platforms.
Yet, it is important to ensure that a synopsis sufficiently represents posts written in different dialects as the dialects are representative of the participating communities.
Studies have shown that the lack of representational diversity can exacerbate negative stereotypes and lead to downstream biases \cite{kay2015unequal, snyder1977social, rickford2016raciolinguistics, terry2010variable}.
Summarization algorithms, in particular, can aggravate negative stereotypes by providing a false perception of the ground truth \cite{kay2015unequal}. Hence, it is crucial for automatically generated text summaries to be dialect-diverse.

\noindent
\subsection{Our Contributions}
We analyze the dialect diversity of standard summarization algorithms that represent the range of paradigms employed for extractive summarization on platforms like Twitter.
This includes frequency-based algorithms (TF-IDF \cite{luhn1957statistical}, Hybrid TF-IDF \cite{inouye2011comparing}), graph algorithms (LexRank \cite{erkan2004lexrank}, TextRank  \cite{mihalcea2004textrank}), algorithms that reduce redundancy (MMR \cite{carbinell2017use}, Centroid-Word2Vec \cite{rossiello2017centroid}), and pre-trained supervised approaches (SummaRuNNer \cite{nallapati2017summarunner}). %
All algorithms use various structural properties of the sentences (Twitter posts, in our case) to score them on their importance.
Our primary evaluation datasets are the TwitterAAE \cite{blodgett2016demographic}, the Crowdflower Gender AI and the Claritin datasets \cite{dash2019summarizing}. 
We observe that, for random and topic-specific collections from the TwitterAAE dataset,
 most algorithms return summaries that under-represent the AAE-dialect.
Similarly, for Crowdflower AI and Claritin datasets, these algorithms often
return gender-imbalanced summaries (Section~\ref{sec:orig_expts}).

To address the dialect bias and utilize the effectiveness of the existing summarization algorithms, we employ a framework that takes any summarization algorithm as a blackbox and returns a summary that is more dialect-diverse than the one the summarization algorithm would return without intervention. 
Along with the blackbox algorithm, this approach needs a small dialect-diverse control set of posts as part of the input; the generated summary is diverse in a similar manner as the control set (Section~\ref{sec:model}).
Importantly, and in contrast to existing work \cite{dash2019summarizing}, by using similarity metrics with items in the control set, the framework bypasses the need for dialect labels in the collection of posts being summarized.
Empirically, we show that our framework improves the dialect diversity of the generated summary for all Twitter datasets and discuss the deviation of the summaries generated by our framework from those generated by the blackbox algorithms and manually-generated summaries (Section~\ref{sec:debias_expts}).
For the Claritin dataset, we also compare the performance against the fair summarization algorithm of \citet{dash2019summarizing}, which explicitly requires labels for diversification.
We observe that the summaries generated by our framework are nearly gender-balanced
and ROUGE scores of these summaries (measuring the similarity between the generated and reference summaries) are close to the ROUGE scores of summaries generated by \citet{dash2019summarizing}.
This comparison further exhibits the effectiveness of using control sets, instead of labels, for diversification.

{
Text summarization on Twitter is useful for search operations; however, there may not be a singular theme associated with the posts being summarized, which makes the context of summarization in this paper slightly different than applications where a single document is summarized into a small paragraph \cite{radev2001experiments}.
In other words, the objective of our paper can be interpreted as data-subsampling with the goal of ensuring content and representational diversity.
}

\noindent
\subsection{Related Work}
\paragraph{Bias in NLP.} Recent studies have explored the presence of social biases in various language processing models. 
Pre-trained encoders \cite{mikolov2015computing, bojanowski2017enriching, devlin2018bert}
have been shown to exhibit gender, racial and intersectional biases \cite{bolukbasi2016man, caliskan2017semantics, tan2019assessing, may2019measuring, nadeem2020stereoset}, 
often leading to social biases in downstream tasks.
This includes gender and racial bias in sentiment-analysis systems \cite{kiritchenko2018examining}, image captioning models \cite{hendricks2018women}, language identification \cite{blodgett2017racial, lu2018gender}, hate/abusive speech detection \cite{sap2019risk, park2018reducing}, and speech recognition \cite{tatman2017gender}.
Considering the significance of these language tasks, techniques to mitigate biases in some of the above NLP applications have been proposed \cite{blodgett2020language, bolukbasi2016man, sun2019mitigating, zhao2017men, zhao2018learning, dash2019summarizing}.
However, dialect diversity in summaries of textual data has not been explicitly considered before and, in the absence of dialect labels, most fair summarization approaches cannot be extended to this problem; our work aims to address both of these issues.

\paragraph{Text summarization algorithms.}
The importance of a sentence in a collection can be quantified in different ways.
Algorithms such as TF-IDF \cite{luhn1957statistical} and Hybrid TF-IDF algorithm \cite{inouye2011comparing} rank sentences based on word and document frequencies.
Other unsupervised algorithms, such as LexRank \cite{erkan2004lexrank}, TextRank \cite{mihalcea2004textrank}, and centroid-based approaches \cite{rossiello2017centroid, miller2019leveraging, padmakumar2016unsupervised}, quantify the importance of a sentence based on how well it represents the collection. 
LexRank and TextRank define a graph over the posts, quantifying the edges using pairwise similarity, and score sentences based on their centrality in the graph.
Along similar lines, \citet{rossiello2017centroid} propose a centroid-based summarization method that uses compositional properties of word embeddings to quantify the similarity between sentences.
To ensure that summary a representative of the collection being summarized, prior algorithms often define \textit{non-redundancy} as a secondary goal \cite{lin2011class}.
This includes Maximum Marginal Relevance score (MMR) \cite{carbinell2017use} algorithm, Maximum Coverage Minimum Redundant (MCMR) models \cite{alguliev2011mcmr}, Determinantal Point Processes \cite{kulesza2012determinantal}, and latent variable based approaches \cite{ozsoy2011text, lee2009automatic}.
The centroid-based approach of \citet{rossiello2017centroid} also has a non-redundancy component.
While adding the sentences with the highest scores to the summary, their algorithm checks for redundancy and if a candidate sentence is \textit{very similar} to a sentence already present in the summary, it is discarded (similar to the greedy MMR approach). %
However, reducing redundancy has been shown to be ineffective in ensuring diversity with respect to specific attributes, such as gender or race, in other applications \cite{celis2016fair, celis2019implicit}.
To empirically demonstrate the ineffectiveness of non-redundancy in ensuring dialect diversity, we analyze the summaries generated by MMR \cite{carbinell2017use} and \citet{rossiello2017centroid} (implemented using Word2Vec embeddings and referred to as \textit{Centroid-Word2Vec} for the rest of the paper) algorithms.

\begin{figure*}[t]
\includegraphics[width=\linewidth]{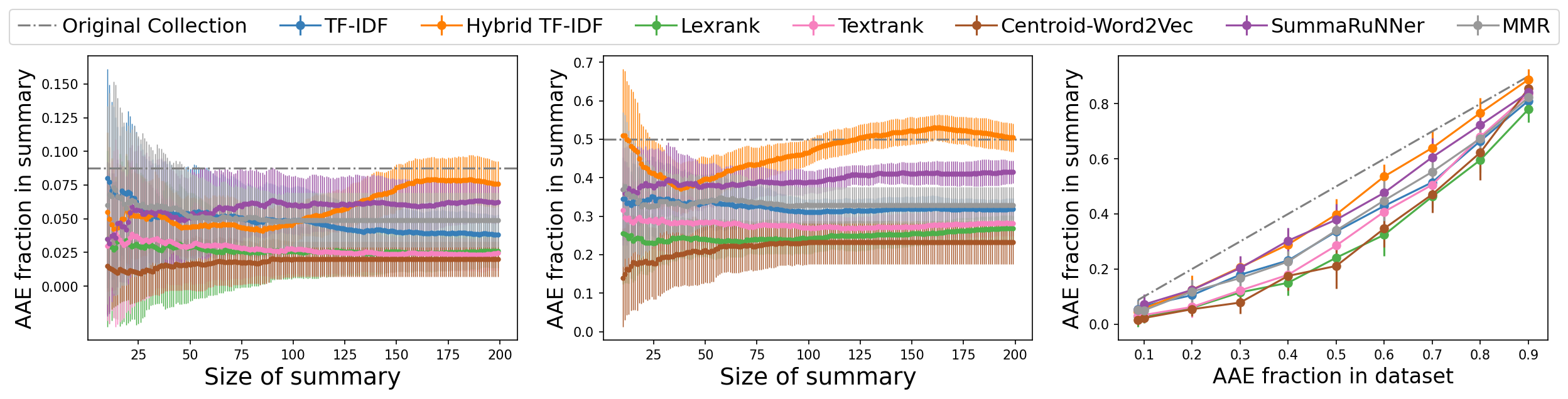}
  \subfloat[Collection has 8.7\% AAE posts]{\hspace{.35\linewidth}}
\subfloat[Collection has 50\% AAE posts]{\hspace{.33\linewidth}}
  \subfloat[Fixed summary size: 50]{\hspace{.31\linewidth}}
\caption{\small\textit{TwitterAAE Evaluation 1.}
Plots (a), (b) presents the dialect diversity of generated summaries when the collection being summarized has 8.7\% and 50\% AAE posts respectively. 
Each point represents to the mean fraction of AAE posts in summary of the given size, with standard error as errorbars. 
Plot (c) presents the dialect diversity in summaries of size 50 vs original collection with varying fraction of AAE posts.
All algorithms other than Hybrid TF-IDF return summaries that have a smaller fraction of AAE posts than the original collection.
}
\label{fig:random_aae_frac}
\end{figure*}

We choose TextRank and Hybrid TF-IDF for our diversity analysis because they have been shown to produce better summaries (evaluated using ROUGE metrics over manually-generated summaries) for Twitter datasets than other frequency, graph, and latent variable based approaches \cite{inouye2011comparing, nguyen2018tsix}.
TF-IDF and LexRank are also commonly used for Twitter datasets and serve as baselines for our analysis.
The original papers for most of these text summarization algorithms focused on evaluation on DUC or CNN/DailyMail datasets; however, the documents in these datasets correspond to news articles that are usually not considerably dialect diverse.

Beyond unsupervised approaches, supervised techniques for summarization classify whether a sentence is important to the summary or not \cite{liu2019text, zhong2020extractive, nallapati2017summarunner, jadhav2018extractive, zhong2019searching}.
These models are trained on datasets for which summaries are available, such as, news articles \cite{hermann2015teaching}, and the models pre-trained on these datasets do not always generalize well to other domains.
We will evaluate the diversity of one such pre-trained model, SummaRuNNer \cite{nallapati2017summarunner}.
\footnote{\textit{Extractive summarization} algorithms use sentences from the collection to create a summary. 
\textit{Abstractive summarization}, on the other hand, aims to capture the semantic information of the dataset and the summary creation can involve paraphrasing the sentences in the dataset \cite{lin2019abstractive}.
Automated diversity evaluation for abstractive summarization algorithms is, therefore, more difficult since the summary is not necessarily a subset of the collection. For this paper, we focus on extractive summarization only.
}
Finally, note that Twitter posts usually have metadata associated with them, and
some algorithms use this metadata to return summaries that are also diverse with respect to time of posts \cite{chellal2018optimization}, and/or user-network \cite{he2018twitter}.
However, since our goal is to analyze the impact of dialect variation on summarization, we focus on techniques that aim to summarize using only the collection of posts.

\paragraph{Prior fair summarization algorithms.}
Most related algorithms that aim to ensure unbiased summarization usually assume the existence of labels or partitions with respect to the socially salient attribute in consideration (in this case, dialect).
For example, \cite{celis2018fair, lin2011class} use labels to construct fairness constraints or scoring functions to guarantee appropriate diversity in automatically generated summaries.
Similarly, for fair text summarization, \citet{dash2019summarizing} propose methods that use socially salient attribute labels to choose representative text summaries for Twitter datasets that are balanced with respect to gender or political leaning of the users.
However, these prior fair summarization approaches are unsuitable for dialect-diverse summarization since dialect labels are not always available (or even desirable \cite{beukeboom2019stereotypes}) for sentence collections encountered in real-world applications and automated dialect classification is a difficult task \cite{jorgensen2015challenges}. 
With the rapidly-evolving nature of dialects on social media, it is unreasonable to rely on existing dialect classification models to obtain accurate dialect labels for every social media post.

Using a dialect-diverse set of examples helps us skirt around the issue of unavailable dialect labels.
The approach of using a diverse control set, instead of labels, to mitigate bias has been previously employed in image-related tasks;
\citet{celis2019implicit} show that a diverse set of example images can be used to improve diversity in image summarization results and \citet{grover2019fair} effectively employ small reference image datasets to obtain unbiased image generative models.
Our framework demonstrates that such small reference sets can be used for fair text summarization as well.

\section{{Dialect diversity of standard text-summarization approaches}} \label{sec:orig_expts}

We examine the dialect diversity of TF-IDF, Hybrid TF-IDF, LexRank, TextRank, Centroid-Word2Vec, MMR and SummaRuNNer.
\footnote{Algorithmic and implementation details of all methods are given in Appendix~\ref{sec:algo_details}.}
All algorithms take as input a collection of Twitter posts and the desired summary size $m$, and return an $m$-sized summary for the collection.

\paragraph{Datasets.}
\textit{TwitterAAE dataset \footnote{\url{http://slanglab.cs.umass.edu/TwitterAAE}}: } Our primary dataset of evaluation is the large TwitterAAE dataset, curated by Blodgett et al. \cite{blodgett2016demographic}.
The dataset overall contains around 60 million Twitter posts from 2013, and for each post, the timestamp, user-id, and geo-location are available as well.
Blodgett et al. \cite{blodgett2016demographic} used the census data 
to learn demographic language models for the following population categories: non-Hispanic Whites, non-Hispanic Blacks, Hispanics, and Asians; using the learned models, they report the probability of each post being written by a user of a given population category.
We pre-process the dataset to filter and remove posts for which the probability of belonging to non-Hispanic African-American English language model or non-Hispanic White English language model is less than 0.99. 
This smaller dataset contains around 102k posts belonging to non-Hispanic African-American English language model and 1.06 million posts belonging to non-Hispanic White English language model;
for simplicity, we will refer to the two groups of posts as AAE and WHE posts in the rest of the paper.

\begin{figure*}[t]
\includegraphics[width=\linewidth]{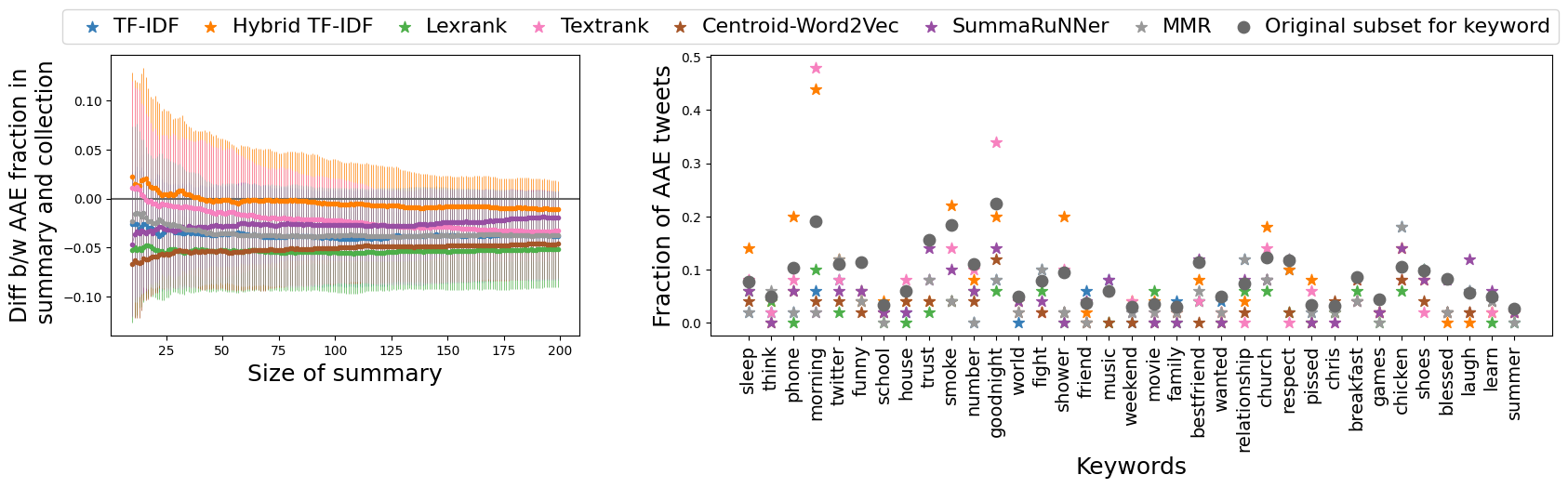}
  \subfloat[Mean dialect diversity vs summary size]{\hspace{.4\linewidth}}
\subfloat[Dialect diversity for different keywords]{\hspace{.6\linewidth}}
\caption{\small\textit{TwitterAAE Evaluation 2.}
Figure (a) reports the mean and standard deviation of the difference between AAE fraction in summary and AAE fraction in the collection of posts that contain the keyword. 
Figure (b) presents fraction of AAE posts in size 50 summaries for different keywords, as well as, the fraction of AAE posts in the subset of posts containing the keyword.
Once again, for most keywords, the algorithms (other than Hybrid TF-IDF) return summaries that have a smaller fraction of AAE posts than the original keyword-specific collection.
}
\label{fig:keyword_aae_frac}
\end{figure*}

We also isolate 35 keywords that occur in a non-trivial fraction of posts in both AAE and WHE partitions to study topic-based summarization\footnote{
Each selected keyword occurs in at least 4500 posts in total and in at least 1500 AAE and WHE posts.}.
The keywords and the fraction of AAE posts in the subset of the dataset containing them are given in Figure~\ref{fig:keyword_aae_frac}.

\noindent
\textit{Claritin Gender dataset \footnote{\url{https://github.com/ad93/FairSumm}}: }
Dialect variation with respect to gender has been received relatively less academic attention;
nevertheless, prior studies have established that there is a recognizable difference between posts by men and posts by women on Twitter \cite{ott2016tweet, miller2012gender}.
Hence, we look at the diversity of summarization algorithms with respect to the fraction of posts by men and women in the generated summaries.
The Claritin dataset contains 3943 Twitter posts about an anti-allergic drug, Claritin,
with 38\% from male user accounts and 62\% from female user accounts.
It was curated to study the possible usage of crowdsourcing to detect gender-specific side-effects and, therefore, we 
look at the diversity of summaries with respect to gender of the account users.
For this dataset, three manually-generated summaries are also available \cite{dash2019summarizing} and will be used to evaluate the utility of our proposed fair summarization framework.

\noindent
\textit{CrowdFlower AI Gender dataset \footnote{\url{https://data.world/crowdflower/gender-classifier-data}}: }
This dataset has around 20,000 posts,
with crowdsourced labels for the gender of the creator of every post (male, female, or brand) and location.
We remove the posts with a location outside US to maintain regional uniformity in the posts.
The filtered dataset contains 6176 posts, with 34\% posts from male user accounts, 35\% posts from female user accounts and the rest are labeled as posts by brands or ``unknown''.

For all datasets, we pre-process the posts to remove URLs, represent all posts in lower-case, replace user mentions with the tag ATMENTION and handle special characters. 
However, we do not remove hashtags since they are, semantically, a part of the posts.

\raggedbottom

\paragraph{Evaluation details.}
Despite the filtering, TwitterAAE dataset is prohibitively large for graph-based algorithms, due to the infeasibility of graph-construction for large datasets.
Hence, we limit our simulations to collections of at most size 5000 and generate summaries of sizes upto 200 for these collections.

\noindent
\textit{TwitterAAE Evaluation 1:}
We sample collections of 5000 posts from the TwitterAAE dataset.
and vary the percentage of AAE posts in the collection from 8.7\% (i.e., percentage of AAE posts in the entire dataset) to 90\%.
Then, we run the standard summarization algorithms for each sampled collection and record the fraction of AAE posts in the generated summaries.
For each fraction, we repeat the process 50 times and report the mean and standard error of the fraction of AAE posts in the generated summaries.

\noindent
\textit{TwitterAAE Evaluation 2:} Next, using the 35 common keywords in this dataset, we extract the collection of posts containing any given keyword.
Once again, we use the summarization algorithms on the extracted collections and report the difference between the fraction of AAE posts in the generated summary and fraction of AAE posts in the collection containing the keyword.
This evaluation aims to assess the dialect diversity of summaries generated for topic-specific set of posts and also lets us verify whether the observations of Evaluation 1 extend to non-random collections.

\noindent
\textit{Claritin Evaluation 1:} 
For the Claritin dataset, since the size is relatively small, we use the summarization algorithms on the entire dataset and report the fraction of posts written by men.

\noindent
\textit{Crowdflower Evaluation 1:} 
For this dataset, we again use the summarization algorithms on the entire dataset and report the fraction of posts written by men (amongst posts written by non-brands).

\begin{remark}
For CrowdFlower AI and Claritin dataset, the evaluation is with respect to the gender of the user who created the post, while for the TwitterAAE dataset, the evaluation is with respect to the dialect label of the post.
The evaluation methods across datasets are different in terms of the attribute used, but the goal is the same, i.e., to assess the \textit{dialect representational diversity} of the generated summaries.
The dialects we consider in this paper are those adopted by social groups and the disparate treatment of these dialects is closely related to the disparate treatment of the groups using these dialects.
While the AAE dialect is not necessarily used only by African-Americans, it is primarily associated with them and studies have shown disparate treatment of AAE dialect can lead to racial bias \cite{rickford2016raciolinguistics,jones2019testifying}.
\end{remark}

\paragraph{Results. }
The results for \textit{TwitterAAE Evaluation 1} are presented in Figure~\ref{fig:random_aae_frac}.
Plots~\ref{fig:random_aae_frac}a, b show that for small summary sizes (less than 200), all algorithms mostly return summaries that have a smaller fraction of AAE posts than the original collection.
For larger summary sizes, summaries generated by Hybrid TF-IDF are relatively more dialect diverse.
Even when the fraction of AAE posts in the original collection is increased beyond 0.5, the fraction of AAE posts in size 50 summaries from all algorithms is less than the fraction of AAE posts in the original collection, as evident from Figure~\ref{fig:random_aae_frac}c.

The results for \textit{TwitterAAE Evaluation 2} are presented in Figure~\ref{fig:keyword_aae_frac}.
For many keywords, the summaries generated by all algorithms have lower dialect diversity than the original collection.
For example, for ``funny'' and ``blessed'', the AAE fraction in summaries generated by all algorithms is less than the AAE fraction in the collection containing the keyword.
There are also keyword-specific collections where the summaries are relatively more diverse; e.g., for keyword ``morning'', summaries generated by Hybrid TF-IDF and TextRank have better dialect diversity (AAE fraction $\geq 0.4$) than the original collection ($0.2$).
However, overall the high variance in Plot~\ref{fig:keyword_aae_frac}a shows that the algorithms are not guaranteed to generate sufficiently diverse summaries for all keywords.

For \textit{Claritin Evaluation 1}, the results are presented in Table~\ref{tbl:claritin_results} (along with results of our ``balanced'' algorithms described in Section~\ref{sec:model}). For this dataset, all standard algorithms generate summaries that are gender-imbalanced (fraction of posts by men either $\geq 0.62$ or $\leq 0.41$).
For \textit{Crowdflower Evaluation 1} (Table~\ref{tbl:crowdflower_results}), TF-IDF, MMR, LexRank, SummaRuNNer return nearly balanced summaries with gender fraction in the range $[0.45, 0.53]$.
However, TextRank, Hybrid-TF-IDF, and Centroid-Word2Vec generate gender-imbalanced summaries (fraction of posts by men $\leq 0.37$).

\paragraph{Discussion. }
The above evaluations demonstrate that none of the standard summarization algorithms consistently generate diverse and unbiased summaries across all datasets.
Dialect-imbalanced original collections is not the sole reason for the dialect bias in the summaries either (as evidenced from Figure~\ref{fig:random_aae_frac}b,c).
A possible reason for the bias is
that the scoring mechanism of all algorithms is affected by structural aspects of the dialect;
e.g., frequency-based algorithms weigh each word in a post by to its frequency.
However, given that vocabulary sizes and average post lengths vary across dialects \cite{blodgett2017racial},
using word frequency to quantify importance can favor one dialect over the other
(see Section~\ref{sec:limitations} for further discussion).

The performance of Centroid-Word2Vec and MMR for Claritin and TwitterAAE also shows that ensuring non-redundancy does necessarily not lead to dialect diversity,
and the lack of diversity of SummaRuNNer summaries demonstrates that pre-trained supervised models do not necessarily generalize to other domains.

Despite the lack of dialect diversity in the generated summaries of these algorithms, prior work have demonstrated their utility  \cite{rossiello2017centroid, radev2001experiments}. 
Hence, it is important to explore ways to exploit the utility of algorithms like Centroid-Word2Vec and, at the same time, ensure that the generated summaries are dialect-diverse.

\begin{algorithm}[t]
\footnotesize
  \caption{Post-processing algorithm for fair summarization} 
\begin{algorithmic}[1]
   \State {\bfseries Input:} Dataset $S$, blackbox algorithm $A$, similarity function $\sim$, diversity control set $T$, parameter $\alpha$, and summary size $m$
   \ForAll{$x, z \in S \times T$}
   	\State $\DS(x,z) \gets (1-\alpha) \cdot A(x) + \alpha \cdot \sim(z, x)$
   	\EndFor
   \State $R \gets \emptyset$
   \While{$|R| < m$}
   		\State $r, \textrm{score} \gets \emptyset$
   		\ForAll{$z \in T$} \Comment{Find posts for each $z$} 
   			\State $y \gets \arg \max_{x \in S} \DS(x,z) $
   			\If{$y \notin r$} \Comment{Checking duplicates} 
	   			\State $r \gets r \cup \set{y}$
	   			\State score$(y) \gets  \DS(x,z)$  \Comment{Scores used for tie-breaks} 	
   				\State $\DS(x, z) \gets -\infty$
	   		\EndIf
   		\EndFor
   		\If{$|R \cup r| \leq m$}  \Comment{If all of $r$ can be added} 
   			\State $R \gets R \cup r$
   		\Else \Comment{Tie-break when $m$ is not a multiple of $r$} 
   			\State $m' \gets m - |r|$ 
   			\State $r' \gets m'$ posts from $r$ with highest score$(y)$
   			\State $R \gets R \cup r'$   			
   		\EndIf
   \EndWhile
   \State  return $R$
\end{algorithmic}
  \label{algo:main}
\end{algorithm}
\raggedbottom

\section{Our model to mitigate dialect bias} \label{sec:model}
We employ a simple framework to correct for the dialect bias in standard summarization algorithms.
Let $S$ denote a collection of sentences. 
Our approach uses any standard summarization algorithm, denoted by $A$, as a blackbox
to return a score $A(x)$, for each $x\in S$. This score represents the importance of sentence $x$ in the collection and we assume that the larger the score, the more important is the sentence.
We also need a function to measure the pairwise similarity between sentences; we will call this function  $\sim(\cdot, \cdot)$.
An example of such a similarity function is presented later.

To implicitly ensure dialect-diversity in the results, we use a \textit{diversity control set} $T$, i.e. a {small} set of sentences that has sufficient representation from each dialect (e.g., an equal number of posts from all relevant dialects).
We return a diverse and relevant summary by appropriately combining the importance score from the blackbox $A$ and the diversity with respect to the control set $T$ in the following manner.
Given a hyper-parameter $\alpha\in[0,1]$, for each $z\in T$, let $\DS: S\times T \rightarrow \R$ denote the following score function:\\
$\DS(x, z) =  (1 - \alpha) \cdot A(x) +  \alpha \cdot \sim(z, x).$
Let $\DS_z$ represent the sorted list $\set{\DS(x, z)}_{x \in S}$ and let $DS_{z,i}$ denote the sentence with the $i$-th largest score in $DS_z$.
Based on these scores, we rank the sentences in $S$ in the following order: first we return sentences that have the largest score for each $z$, i.e.,  $\set{DS_{z,1}}_{z \in T}$.
Next, we return the set $\set{DS_{z,2}}_{z \in T}$ and so on.
Sentences within each set $\set{DS_{z,i}}_{z \in T}$ can be ranked by their scores from algorithm $A$. 
At every step, for each $z$ we check if a sentence has already been ranked; if so, we replace it with the sentence with next-highest score for that $z$, ensuring that duplicates are not processed.
The summary based on this ranking can then be generated.
By giving equal importance to every post in $T$ in the ranking, our framework tries to generate a summary that is diverse in a similar manner as $T$.
The complete implementation is provided in Algorithm~\ref{algo:main}.
We will refer to our algorithm, with blackbox $A$ and $\alpha = 0.5$, as $A$-balanced. For example, our algorithm with $A$ as Centroid-Word2Vec will be called \textit{Centroid-Word2Vec-balanced}.

\begin{table}[t]
\centering
\caption{\small \textit{Claritin Evaluation 1.} 
We report the gender diversity and average ROUGE scores of generated summaries (size 100) against the three manually-generated summaries.
For all blackbox algorithms $A$, our post-processed algorithm $A$-balanced returns more gender-balanced summaries than $A$ (marked by \greencircle{1.5pt}).}
\label{tbl:claritin_results}
\footnotesize
\begin{tabular}{@{}lcllll@{}}
\toprule
\multirow{2}{*}{Method} & \multirow{2}{*}{\makecell{\% of posts\\ by men in\\ summary}} & \multicolumn{2}{c}{ ROUGE-1} & \multicolumn{2}{c}{ ROUGE-L} \\ 
\cmidrule(r{1pt}){3-6}
&  & {Recall} & {F-score}  & {Recall} & {F-score}  \rule{0pt}{3ex}    \\
\midrule
Original collection & 0.38 \whitecircle{1.5pt}  &  - &  - &  - & - \\
FairSumm & {0.50} \whitecircle{1.5pt}  & 0.57 & {0.53} & 0.30 & 0.33\\ 
MMR & 0.30 \whitecircle{1.5pt}  & 0.48 & 0.31 & 0.35 & 0.27\\ 
\midrule
TF-IDF & 0.31 \whitecircle{1.5pt} & 0.62 & 0.40 & 0.40 & 0.28 \\
TF-IDF-balanced & 0.35 \greencircle{1.5pt} & 0.63 & 0.44 & 0.40 & 0.30 \\
\hdashline
Hybrid TF-IDF & 0.62 \whitecircle{1.5pt}  & 0.23 & 0.27 & 0.11 & 0.16 \\
Hybrid TF-IDF-balanced & 0.54 \greencircle{1.5pt} & 0.32 & 0.32 & 0.18 & 0.22 \\
\hdashline
Lexrank & 0.41 \whitecircle{1.5pt}  & 0.54 & 0.40 & 0.32 & 0.28 \\
Lexrank-balanced & {0.50} \greencircle{1.5pt} & 0.50 & 0.44 & 0.32 & 0.30 \\
\hdashline
Textrank & 0.62 \whitecircle{1.5pt}  & 0.22 & 0.24 & 0.09 & 0.14 \\
Textrank-balanced & 0.52 \greencircle{1.5pt} & 0.33 & 0.33 & 0.19 & 0.23 \\
\hdashline
SummaRuNNer & 0.35 \whitecircle{1.5pt}  & 0.62 & 0.49 & 0.42 & 0.32 \\
SummaRuNNer-balanced & 0.43 \greencircle{1.5pt} & 0.56 & 0.45 & 0.38 & 0.32 \\
\hdashline
Centroid-Word2Vec & 0.41 \whitecircle{1.5pt}  & 0.61 & 0.44 & 0.38 & 0.33 \\
Centroid-Word2Vec-balanced & 0.44 \greencircle{1.5pt} & {0.58} & 0.45 & {0.36} & {0.33} \\
\bottomrule
\end{tabular}
\end{table}

The idea of summarization based on a linear combination of scores that correspond to different goals has been used in other contexts.
For topic-focused summarization, Vanderwende et al. \cite{vanderwende2007beyond} score each word by linearly adding its frequency and topic relevance score. 
Even MMR computes a linear combination of the importance and non-redundancy score, measured as the maximum similarity to an existing summary sentence.
As mentioned earlier, our approach is based on the fair image summarization approach of \cite{celis2019implicit} that uses diverse examples to generate a diverse image summary.

\noindent
\paragraph{Time complexity.} Let $\mathcal{T}_S$ denote the time taken by blackbox algorithm $A$ to score all elements of $S$. 
To create the $\DS$ matrix, there will be an additive factor of $|T| \times |S|$.
Selecting the best element in each $\DS_z$ can be done in two ways, i.e., either by sorting each $\DS_z$ or using a max-heap over each $\DS_z$.
In both cases, the overall time complexity is $\mathcal{T}_S + (|T| +m) \cdot |S| \cdot \log |S| $.

\noindent
\paragraph{Choice of diversity control sets.}
As mentioned earlier, a diversity control set in our framework 
is used to ensure that generated summary has sufficient representation from every dialect.
Considering the importance of the diversity control set to our framework, the appropriate construction of such sets deserves necessary attention.

We provide one formal mechanism to construct such diversity control sets.
Suppose we have a small set of dialect-labelled posts $V$ (e.g., obtained via human annotation or crowdsourcing).
To construct a control set from $V$, we can extract a smaller subset $T$ (with equal number of posts from all dialects) of $V$ and measure how well it can predict the dialect labels of the posts in $V \setminus T$; here, the predicted label for any post $x \in V \setminus T$ is the dialect label of the post in $T$ with which $x$ has the highest pairwise similarity.
The chosen diversity control set $T$ is the subset with the best prediction score.

For the TwitterAAE dataset, such a $V$ (with human-annotated dialect labels) exists \cite{blodgett2018twitter} with $|V| = 500$.
Since the time complexity of the algorithm depends linearly on the size of this set, we use
the above process to select a diversity control set $T$ of size 28 for our empirical evaluation (see Appendix~\ref{sec:div_control_analysis}).
Note that this is one way of constructing diversity control sets and,
in general, the control set will be context-dependent; they can be hand chosen as well and we discuss the nuances of the composition further in Section~\ref{sec:limitations}.

\begin{table}[t]
\centering
\caption{\small\textit{Crowdflower Evaluation 1.} 
We report the gender diversity (fraction of non-brand posts by male user accounts) and ROUGE scores of $A$-balanced summaries against the summaries generated by $A$, for all $A$ (summary size 100).
Settings where $A$-balanced generates more/equally dialect-diverse summaries than $A$ are marked with \greencircle{1.5pt} and settings where $A$-balanced is worse are marked with \textcolor{red}{$\star$}.
}
\label{tbl:crowdflower_results}
\footnotesize
\begin{tabular}{@{}lcllll@{}}
\toprule
\multirow{2}{*}{Method} & \multirow{2}{*}{\makecell{\% of non-brand\\posts by men\\in summary}} & \multicolumn{2}{c}{ ROUGE-1} & \multicolumn{2}{c}{ ROUGE-L} \\ 
\cmidrule(r{1pt}){3-6}
&  & \scriptsize{Recall} & \scriptsize{F-score}  & \scriptsize{Recall} & \scriptsize{F-score}  \rule{0pt}{3ex}    \\
\midrule
Original collection & 0.49 \whitecircle{1.5pt} & - &  - &  - &  - \\
 MMR & 0.45 \whitecircle{1.5pt}  & - & - & - & -\\ 
\midrule
TF-IDF & 0.53 \whitecircle{1.5pt} & - & - & - & - \\
TF-IDF-balanced & 0.44\textcolor{red}{$\star$} & 0.70 & 0.71 & 0.68 & 0.64 \\
\hdashline
Hybrid TF-IDF & 0.35 \whitecircle{1.5pt} & - & - & - & - \\
Hybrid TF-IDF-balanced & 0.40 \greencircle{1.5pt} & 0.84 & 0.63 & 0.61 & 0.46 \\
\hdashline
Lexrank & 0.46 \whitecircle{1.5pt} & - & - & - & - \\
Lexrank-balanced & 0.47 \greencircle{1.5pt} & 0.59 & 0.59 & 0.43 & 0.40 \\
\hdashline
Textrank & 0.37 \whitecircle{1.5pt} & - & - & - & - \\
Textrank-balanced & 0.34\textcolor{red}{$\star$} & 0.82 & 0.81 & 0.78 & 0.73 \\
\hdashline
SummaRuNNer & 0.50 \whitecircle{1.5pt} & - & - & - & - \\
SummaRuNNer-balanced & 0.50 \greencircle{1.5pt} & 0.76 & 0.73 & 0.66 & 0.68 \\
\hdashline
Centroid-Word2Vec & 0.34 \whitecircle{1.5pt} & - & - & - & - \\
Centroid-Word2Vec-balanced & 0.40 \greencircle{1.5pt} & 0.70 & 0.70 & 0.54 & 0.51 \\
\bottomrule
\end{tabular}
\end{table}
\raggedbottom

\section{Empirical analysis of our model} \label{sec:debias_expts}
We repeat the evaluations proposed in Section~\ref{sec:orig_expts} for our post-processing framework, i.e., \textit{TwitterAAE Evaluation 1 \& 2, CrowdFlower Evaluation 1, and Claritin Evaluation 1}.
For the Claritin dataset, we also compare against the FairSumm algorithm of \citet{dash2019summarizing}; FairSumm explicitly requires access to dialect labels and comparison against this baseline lets us assess the performance of our framework, that uses diversity control sets for diversification, to an algorithm that uses attribute labels for diversification.
For this dataset, \citet{dash2019summarizing} provide three manually-generated summaries of size 100 and we evaluate the summaries generated by all algorithms according to average similarity with the manually-generated summaries.
The measure of evaluation employed is ROUGE recall and F-scores \cite{lin2003automatic}.
To state briefly, ROUGE-1 scores quantify the amount of unigram overlap between the generated summary and the reference summary, and ROUGE-L scores looks at the longest co-occurring sequence in the generated and reference summary
\footnote{The best average ROUGE-1 recall and F-score achieved for Claritin dataset (against the three manually-generated reference summaries), by any algorithm considered in this paper or \cite{dash2019summarizing}, is 0.62 and 0.57 respectively.}.
For the other datasets, since we do not have manually-generated summaries, we
use ROUGE scores to compare against summaries from the standard summarization algorithms.

The diversity control set chosen for TwitterAAE evaluations contains 28 posts, with an equal number of AAE and WHE posts, and the sets used for Crowdflower and Claritin evaluations contain 40 and 20 posts respectively, with an equal number of posts written by male and female user accounts.
Details of these sets are provided in Appendix~\ref{sec:div_control_analysis}.

We use the following similarity function for a given pair of sentences $x_1, x_2$: $\sim(x_1, x_2) := 1 - \textrm{cosine-distance}(v_{x_1}, v_{x_2})$, where $v_x$ denotes the feature vector of sentence $x$.
To obtain feature vectors for the sentences, we use a publicly-available word2vec model pre-trained on a corpus of 400 million Twitter posts \cite{godin2019}.
First, we use the word2vec model to get feature vectors for the words in a sentence and then
aggregate them by computing a weighted average, where the weight assigned to a word is proportional to the smooth inverse frequency of the word (see Arora et al. \cite{arora2016simple}).

\begin{figure*}[t]
\small
\includegraphics[width=\linewidth]{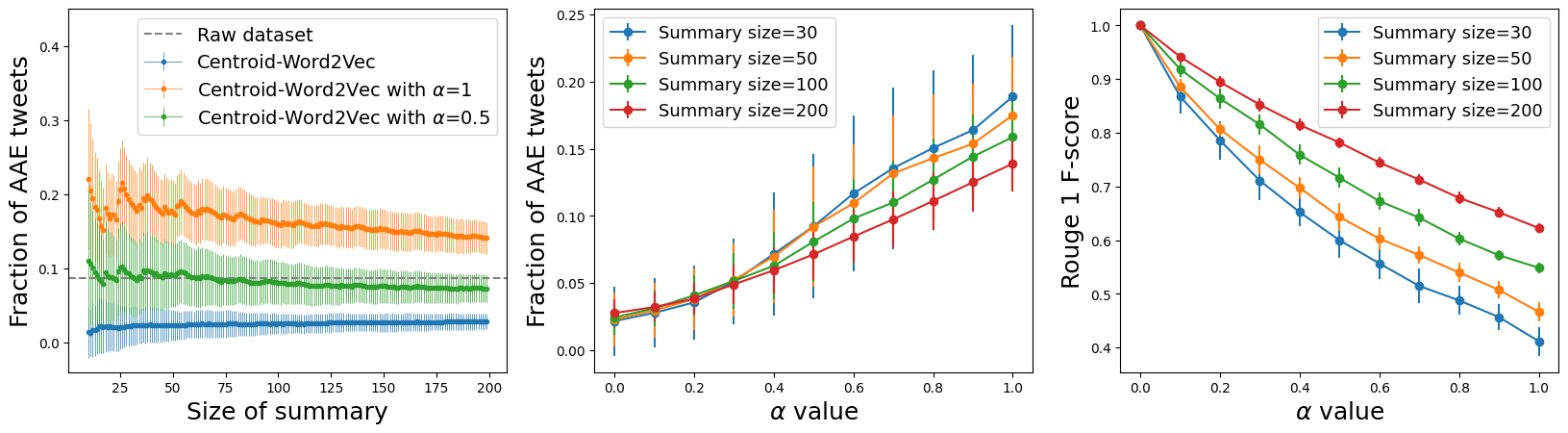}
  \subfloat[AAE fraction vs summary size]{\hspace{.40\linewidth}}
\subfloat[AAE fraction vs $\alpha$]{\hspace{.23\linewidth}}
  \subfloat[Rouge-1 F-score vs $\alpha$]{\hspace{.37\linewidth}} \\
\includegraphics[width=\linewidth]{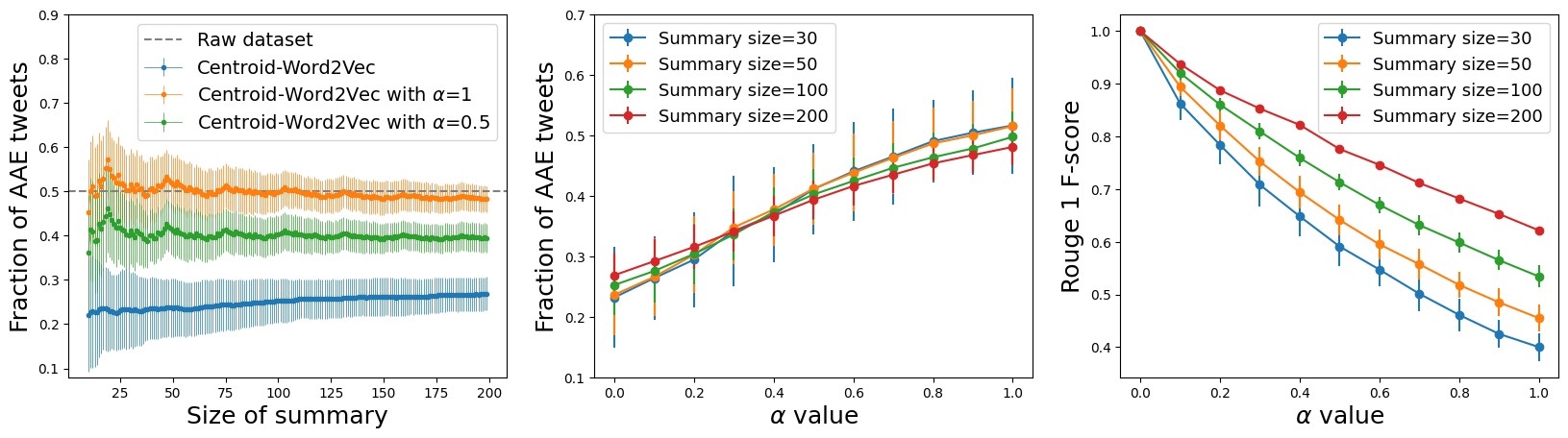}
  \subfloat[AAE fraction vs summary size]{\hspace{.40\linewidth}}
\subfloat[AAE fraction vs $\alpha$]{\hspace{.23\linewidth}}
  \subfloat[Rouge-1 F-score vs $\alpha$]{\hspace{.37\linewidth}} 
\caption{\small 
The first and second row present the evaluation of Centroid-Word2Vec-balanced on collections containing 8.7\% and 50\% AAE posts respectively.
Plots (a), (d) present the fraction of AAE posts for different summary sizes.
Plots (b), (e) present the diversity variation with $\alpha$ and plots (c), (f) present ROUGE-1 F-score between summaries generated using Centroid-Word2Vec-balanced and Centroid-Word2Vec.
For both settings, Centroid-Word2Vec-balanced generates summaries that are significantly more diverse than Centroid-Word2Vec.
}
\label{fig:random_aae_frac_fair}
\end{figure*}
\raggedbottom

\noindent
\paragraph{Results.}
The performance of our framework for \textit{Claritin Evaluation 1} is presented in Table~\ref{tbl:claritin_results}.
We can quantify the gender balance of a summary as the deviation of the fraction of posts by men in the summary from 0.50.
For all algorithms $A$, our framework $A$-balanced generates summaries that are more gender-balanced than summaries of $A$.
Infact, the fraction of posts by men in the summaries generated by the balanced versions of all algorithms, other than TF-IDF, is in the range $[0.43,0.54]$.
Baseline FairSumm (which requires dialect labels), as expected, returns a gender-balanced summary.
ROUGE evaluation with respect to manually-generated summaries also shows that the loss in utility for some balanced algorithms, as compared to the summary generated by FairSumm, is not large. 
The average ROUGE-1 recall of Centroid-Word2Vec-balanced and SummaRuNNer-balanced summaries, with respect to the three reference summaries, is 0.56 and 0.58 respectively; in comparison, the average ROUGE-1 recall of the summary generated by FairSumm is 0.57; however, the precision of Centroid-Word2Vec-balanced and SummaRuNNer-balanced summaries is slightly lower, resulting in a lower ROUGE-1 F-score compared to FairSumm summary.
With respect to ROUGE-L, Centroid-Word2Vec-balanced summary has better recall  and same F-score (0.36 and 0.33) as FairSumm summary (0.30 and 0.33).
The results show that even without access to gender labels, our framework returns nearly gender-balanced summaries, whose utility (as measured using ROUGE evaluation with reference summaries) is comparable to that of FairSumm summary, which explicitly needs gender labels for diversification.
Interestingly, for Hybrid TF-IDF and TextRank which have low initial ROUGE-1 recall ($\leq 0.23$) and F-scores ($\leq 0.27$), using our post-processing framework helps improve these utility scores by forcing the selection of a diverse set of posts.
{Additional manual comparison 
shows that reference summaries, on average, had 63 relevant posts (i.e., posts about usage or side-effects of the drug), while the summary generated by Centroid-Word2Vec-balanced had 56 relevant posts. In this context, the summary generated by our algorithm is more dialect-diverse but suffers a minimal decrease in utility.
}

The performance for \textit{Crowdflower Evaluation 1} is presented in Table~\ref{tbl:crowdflower_results}. 
Once again, the summary generated by Centroid-Word2Vec-balanced is more balanced; the fraction of non-brand posts by men in Centroid-Word2Vec-balanced summary is 0.40, whereas it is 0.34 in Centroid-Word2Vec summary.
Similarly, for Summa-RuNNer-balanced, LexRank-balanced, and TF-IDF-balanced, the fraction of posts by men in the generated summaries is in range $[0.44, 0.50]$.
However, TF-IDF-balanced and TextRank-balanced return relatively less gender-balanced summaries than their blackbox counterparts; in this case, better diversity in the summary can be achieved by using a larger $\alpha$ value or a different control set.
The results using different $\alpha$ values and summary sizes are presented in Appendix~\ref{sec:gender_appendix}.

For \textit{TwitterAAE Evaluation 1}, the detailed performance of our model
using Centroid-Word2Vec as the blackbox algorithm, is presented in Figure~\ref{fig:random_aae_frac_fair}.
Plots~\ref{fig:random_aae_frac_fair}a,d show that using our model with $\alpha = 0.5$ (Centroid-Word2Vec-balanced) leads to improved dialect diversity in the summary (statistically different AAE fraction means). 
For the case when the initial collection has 50\% AAE posts, Centroid-Word2Vec-balanced generates summaries that have 40\% AAE posts in the summary; to achieve better dialect diversity in summary, $\alpha$ value needs to be increased (Plot~\ref{fig:random_aae_frac_fair}e).
The detailed performance on \textit{TwitterAAE Evaluation 2} for two keywords, ``twitter'' and ``funny'', is presented in Table~\ref{tbl:results_query_fair}.
We see that our framework leads to a higher fraction of AAE posts in summary in most cases, compared to just the blackbox algorithm.
However, it does not always improve the diversity;
eg, for keyword ``funny'' and TextRank as the blackbox, the fraction of AAE posts in summaries from the balanced version (0.04) is less than that from just the blackbox (0.06).
In this case, either $\alpha$ or the fraction of AAE posts in the control set can be made larger to generate a more diverse summary.
%
See Appendix~\ref{sec:aae_other} for performance using different keywords, blackbox algorithms, and $\alpha$.

\begin{table*}[t]
\setlength{\tabcolsep}{3pt}
\caption{\small \textit{TwitterAAE Evaluation 2.} The performance of our framework for keywords ``twitter'' and ``funny''.
The ROUGE scores are computed for $A$-balanced summaries against summaries generated by $A$ (summary size 50).
Settings where $A$-balanced summary has a larger fraction of AAE posts than $A$ are marked with \greencircle{1.5pt} and settings where $A$-balanced has a smaller fraction are marked with \textcolor{red}{$\star$}. For all but three settings, $A$-balanced returns summaries with a larger fraction of AAE posts than $A$, at the cost of certain deviation from the summaries of $A$.
}
\label{tbl:results_query_fair}
\footnotesize
\begin{tabular}{@{}l|ccccc|ccccc@{}}
\toprule
& \multicolumn{5}{c|}{ Keyword: ``twitter''} & \multicolumn{5}{c}{ Keyword: ``funny''} \\ \rule{0pt}{3ex}    
\multirow{2}{*}{Method} & \multirow{2}{*}{\makecell{\% AAE in\\ summary}} & \multicolumn{2}{c}{ ROUGE-1} & \multicolumn{2}{c|}{ ROUGE-L} & \multirow{2}{*}{\makecell{\% AAE in\\ summary\\ }} & \multicolumn{2}{c}{ ROUGE-1} & \multicolumn{2}{c}{ ROUGE-L} \\
\cmidrule(r{1pt}){3-6}
\cmidrule(r{1pt}){8-11} 

 &  & {Recall} & {F-score}  & {Recall} & {F-score} &  & {Recall} & {F-score}  & {Recall} & {F-score} \\
 \midrule
 {Collection with keyword}  & 0.11 \whitecircle{1.5pt}  & - & -  & - & -  & 0.10 \whitecircle{1.5pt}  & - & -  & - & - \\
 \hdashline
  TF-IDF & 0.10 \whitecircle{1.5pt} & - &  - & - & -  &0.04 \whitecircle{1.5pt} & -  & - & - & -\\
  TF-IDF-balanced & 0.16 \greencircle{1.5pt} & 0.72 & 0.74 & 0.71 & 0.70 & 0.10 \greencircle{1.5pt} & 0.76 & 0.78 & 0.77 & 0.75  \\
 \hdashline
   Hybrid-TF-IDF & 0.08 \whitecircle{1.5pt} &  - & - & - & - & 0.04 \whitecircle{1.5pt} &  - & - & - & -  \\
   {Hybrid-TF-IDF-balanced} & 0.10 \greencircle{1.5pt} & 0.85 & 0.59 & 0.69 & 0.45 & 0.04\textcolor{red}{$\star$} & 0.89 &  0.54 & 0.78 & 0.33  \\
 \hdashline
   LexRank & 0.04 \whitecircle{1.5pt} & - & - & - & -  & 0.04 \whitecircle{1.5pt} & -  & - & - & -  \\
  LexRank-balanced  & 0.22 \greencircle{1.5pt} & 0.49 & 0.51 & 0.33 & 0.30 & 0.22 \greencircle{1.5pt} & 0.53 & 0.54 & 0.41  & 0.38  \\
 \hdashline
   TextRank & 0.09 \whitecircle{1.5pt} & - & - & - & - & 0.06 \whitecircle{1.5pt} & - & - & - & - \\
  TextRank-balanced & 0.06\textcolor{red}{$\star$} & 0.96 & 0.76 & 0.93 & 0.73 & 0.04\textcolor{red}{$\star$} & 0.94 & 0.43 & 0.92 & 0.25   \\
 \hdashline
   SummRuNNer & 0.08 \whitecircle{1.5pt} & - & - & - & - & 0.06 \whitecircle{1.5pt} & - & - & - & -  \\
  SummRuNNer-balanced & 0.16 \greencircle{1.5pt} & 0.57 & 0.55 & 0.42 & 0.40 & 0.12  \greencircle{1.5pt} & 0.75 & 0.69 & 0.68 & 0.64   \\
 \hdashline
  Centroid-Word2Vec & 0.06 \whitecircle{1.5pt} & - & - & - & - & 0.02 \whitecircle{1.5pt} & - & - & - & -   \\
  Centroid-Word2Vec-balanced & 0.12 \greencircle{1.5pt} & 0.64 & 0.65 & 0.51 & 0.47  & 0.10 \greencircle{1.5pt} & 0.68 & 0.67 & 0.57 & 0.53 \\
\bottomrule
\end{tabular}
\end{table*}

The ROUGE scores for \textit{TwitterAAE Evaluation 1} are presented in Figure~\ref{fig:random_aae_frac_fair}c, f.
As expected, the similarity between the summary generated by our model and summary generated by Centroid-Word2vec decreases as the $\alpha$ increases.
For summary-size 200, the ROUGE-1 F-score is greater than 0.7, implying significant word-overlap between the two summaries.
ROUGE scores in Table~\ref{tbl:results_query_fair} shows that, for \textit{TwitterAAE Evaluation 2}, if the diversity correction required is small, then the recall scores tend to be large.  For Centroid-Word2Vec-balanced, the recall is greater than 0.64, implying that the Centroid-Word2Vec-balanced summary covers at least 64\% of the words in the summary of the blackbox algorithm. 
However, in the cases when the summaries generated by blackbox algorithm originally have low dialect diversity, the recall scores tend to be small (e.g., LexRank-balanced has recall around 0.5).
In these cases, a larger deviation from the original summaries is necessary to ensure sufficient dialect diversity.
With respect to the ROUGE-assessment for TwitterAAE evaluations, note that this measure does not necessarily quantify the usability or the accuracy of the summaries in this case; it simply looks at the amount of deviation from summaries of the blackbox algorithms.

\section{{Discussion, Limitations \& Future work}} \label{sec:limitations}
Our post-processing framework provides a simple mechanism that uses standard summarization algorithms to generate diverse summaries. 
Yet, there are computational and societal aspects along which the framework can be further analyzed.

\noindent
\paragraph{Analyzing the source of dialct bias.} 
While we present empirical evidence that the standard summarization algorithms often generate dialect-biased summaries, it is critical to further delve into the source of such bias.
An important empirical observation was that, for TwitterAAE evaluations, Hybrid TF-IDF generated relatively more dialect-balanced summaries than other algorithms but did not generate dialect-balanced summaries for CrowdFlower evaluation. 
Similarly, TF-IDF generated balanced summaries for CrowdFlower, but not for other evaluations.
As mentioned earlier, this performance discrepancy of the algorithms across datasets is likely related to the design of the algorithms and the structural aspects of the posts they use to generate summaries.
There are often structural differences between sentences written in different dialects.
For instance, an AAE post contains around 8 words on average, while a WHE post contains around 11 words on average. The vocabulary size of all AAE posts in the TwitterAAE dataset is around 57k, while for WHE posts it is around 258k.
We believe that these structural differences lead to the algorithms treating the dialects differently, resulting in dialect-imbalanced summaries.
While we limit our analysis to empirical dialect diversity evaluation, future work on this topic can explore the underlying causes for the dialect bias and suggest possible improvements to the standard summarization algorithms that directly address this bias.

\noindent
\paragraph{Diversity control sets.}
While we provide an automated mechanism to construct diversity control sets (Appendix~\ref{sec:div_control_analysis}), there are limitations to using this construction method.
It crucially uses the dialect partitions in the smaller labelled dataset to construct the control set and, as discussed before, these partitions may not be desirable or capture the evolving nature of dialects.
To mitigate this, the diversity control sets need to be regularly updated to include posts that better reflect the dialects of the user-base.

In general, the choice of diversity control set is context-dependent, and the societal and policy impact of the control set composition requires careful deliberation.
Dialects represent communities and the boundaries between dialects are quite fluid \cite{eisenstein2010latent}.
Correspondingly, deciding whether a control set sufficiently represents any specific dialect or not can be better answered by a person who writes in that dialect than a automated classification/clustering model which constantly needs a large number of diverse sentences for training.
Hence, another way to ensure that the composition of the diversity control set has sufficient representation from all user-dialects is to get the feedback of the communities representing the user-base of the application. 
This would involve 
regular public audits and mechanisms to incorporate community assessment on the control set composition.
Having a {small} and interpretable control set (as in our case) makes this process less cumbersome.
%
Further, by incorporating community feedback into the design of control sets, our framework lets users have a say in the representational diversity of the summaries.
Such participatory designs lead to more cooperative frameworks and are encouraged in fairness literature \cite{sassaman2020creating,chancellor2019relationships}.

Finally, note that using a misrepresentative control set can lead to less diverse summaries; e.g., using sentences in the control set that represent a different set of dialects than the dataset can lead to a worse summary.
To prevent this, the fairness-utility tradeoff
should be taken into account while deciding the control set composition.

\noindent
\paragraph{Improved implementation.}
Depending on the application, the choice of pre-trained embeddings and similarity function can be varied to improve the performance.
For example, instead of using cosine distance of aggregated features of all the words in a given post, one could identify words that differ across dialects and measure similarity with respect to these words only.
It is also important to note that there are issues associated with ROUGE evaluations of generated summaries, such as lack of emphasis on factual correctness \cite{kryscinski2019neural}. 
Recent work has proposed summary generation methods that are factually consistent \cite{celikyilmaz2020evaluation} and extensions of our post-processing framework for such methods can explored as part of future work.

\noindent
\paragraph{Other domains.}
Another important future direction is to inspect diversity of the algorithms for domains beyond Twitter, with sentences written in other languages,
and methods to evaluate the diversity of summaries from abstractive summarization algorithms.

\section{Conclusion}
This paper addresses the issue of dialect diversity in automatically generated summaries for Twitter datasets.
We show that standard summarization algorithms often return summaries that are dialect biased and
employ a framework that uses a small set of dialect diverse posts to improve the diversity of generated summaries.
By using standard summarization algorithms as blackbox, we seek to exploit their utility and by using sets of diverse examples, we ensure that the framework is independent of dialect classification tools.

\section*{Acknowledgements}
This research was supported in part by a J.P. Morgan Faculty Award.  We would like to thank Kush Varshney for early discussions on this problem.

\bibliographystyle{plainnat}
\bibliography{references}

\begin{thebibliography}{72}
\providecommand{\natexlab}[1]{#1}
\providecommand{\url}[1]{\texttt{#1}}
\expandafter\ifx\csname urlstyle\endcsname\relax
  \providecommand{\doi}[1]{doi: #1}\else
  \providecommand{\doi}{doi: \begingroup \urlstyle{rm}\Url}\fi

\bibitem[Alguliev et~al.(2011)Alguliev, Aliguliyev, Hajirahimova, and
  Mehdiyev]{alguliev2011mcmr}
Rasim~M Alguliev, Ramiz~M Aliguliyev, Makrufa~S Hajirahimova, and Chingiz~A
  Mehdiyev.
\newblock Mcmr: Maximum coverage and minimum redundant text summarization
  model.
\newblock \emph{Expert Systems with Applications}, 38\penalty0 (12), 2011.

\bibitem[Arora et~al.(2017)Arora, Liang, and Ma]{arora2016simple}
Sanjeev Arora, Yingyu Liang, and Tengyu Ma.
\newblock A simple but tough-to-beat baseline for sentence embeddings.
\newblock In \emph{ICLR 2017}, 2017.

\bibitem[Beukeboom and Burgers(2019)]{beukeboom2019stereotypes}
Camiel~J Beukeboom and Christian Burgers.
\newblock How stereotypes are shared through language: a review and
  introduction of the aocial categories and stereotypes communication
  framework.
\newblock \emph{Review of Communication Research}, 2019.

\bibitem[Blodgett and O'Connor(2017)]{blodgett2017racial}
Su~Lin Blodgett and Brendan O'Connor.
\newblock Racial disparity in natural language processing: A case study of
  social media african-american english.
\newblock 2017.

\bibitem[Blodgett et~al.(2016)Blodgett, Green, and
  O’Connor]{blodgett2016demographic}
Su~Lin Blodgett, Lisa Green, and Brendan O’Connor.
\newblock Demographic dialectal variation in social media: A case study of
  african-american english.
\newblock In \emph{2016 Conference on Empirical Methods in Natural Language
  Processing}, 2016.

\bibitem[Blodgett et~al.(2018)Blodgett, Wei, and
  O’Connor]{blodgett2018twitter}
Su~Lin Blodgett, Johnny Wei, and Brendan O’Connor.
\newblock Twitter universal dependency parsing for african-american and
  mainstream american english.
\newblock In \emph{Proceedings of the 56th Annual Meeting of the Association
  for Computational Linguistics (Volume 1: Long Papers)}, pages 1415--1425,
  2018.

\bibitem[Blodgett et~al.(2020)Blodgett, Barocas, Daum\'e, and
  Wallach]{blodgett2020language}
Su~Lin Blodgett, Solon Barocas, Hal Daum\'e, III, and Hanna Wallach.
\newblock Language (technology) is power: A critical survey of ``bias'' in nlp.
\newblock In \emph{Proceedings of the Conference of the Association for
  Computational Linguistics (ACL)}, 2020.

\bibitem[Bojanowski et~al.(2017)Bojanowski, Grave, Joulin, and
  Mikolov]{bojanowski2017enriching}
Piotr Bojanowski, Edouard Grave, Armand Joulin, and Tomas Mikolov.
\newblock Enriching word vectors with subword information.
\newblock \emph{Transactions of the Association for Computational Linguistics},
  5:\penalty0 135--146, 2017.
\newblock ISSN 2307-387X.

\bibitem[Bolukbasi et~al.(2016)Bolukbasi, Chang, Zou, Saligrama, and
  Kalai]{bolukbasi2016man}
Tolga Bolukbasi, Kai-Wei Chang, James~Y Zou, Venkatesh Saligrama, and Adam~T
  Kalai.
\newblock Man is to computer programmer as woman is to homemaker? debiasing
  word embeddings.
\newblock In \emph{Advances in neural information processing systems}, 2016.

\bibitem[Caliskan et~al.(2017)Caliskan, Bryson, and
  Narayanan]{caliskan2017semantics}
Aylin Caliskan, Joanna~J Bryson, and Arvind Narayanan.
\newblock Semantics derived automatically from language corpora contain
  human-like biases.
\newblock \emph{Science}, 2017.

\bibitem[Celikyilmaz et~al.(2020)Celikyilmaz, Clark, and
  Gao]{celikyilmaz2020evaluation}
Asli Celikyilmaz, Elizabeth Clark, and Jianfeng Gao.
\newblock Evaluation of text generation: A survey.
\newblock \emph{arXiv preprint arXiv:2006.14799}, 2020.

\bibitem[Celis et~al.(2018)Celis, Keswani, Straszak, Deshpande, Kathuria, and
  Vishnoi]{celis2018fair}
Elisa Celis, Vijay Keswani, Damian Straszak, Amit Deshpande, Tarun Kathuria,
  and Nisheeth Vishnoi.
\newblock Fair and diverse dpp-based data summarization.
\newblock In \emph{International Conference on Machine Learning}, 2018.

\bibitem[Celis and Keswani(2020)]{celis2019implicit}
L~Elisa Celis and Vijay Keswani.
\newblock Implicit diversity in image summarization.
\newblock In \emph{ACM CSCW 2020}, 2020.

\bibitem[Celis et~al.(2016)Celis, Deshpande, Kathuria, and
  Vishnoi]{celis2016fair}
L~Elisa Celis, Amit Deshpande, Tarun Kathuria, and Nisheeth~K Vishnoi.
\newblock How to be fair and diverse?
\newblock \emph{arXiv preprint arXiv:1610.07183}, 2016.

\bibitem[Chancellor et~al.(2019)Chancellor, Guha, Kaye, King, Salehi,
  Schoenebeck, and Stowell]{chancellor2019relationships}
Stevie Chancellor, Shion Guha, Jofish Kaye, Jen King, Niloufar Salehi, Sarita
  Schoenebeck, and Elizabeth Stowell.
\newblock The relationships between data, power, and justice in cscw research.
\newblock In \emph{Conference Companion Publication of the 2019 on Computer
  Supported Cooperative Work and Social Computing}, pages 102--105, 2019.

\bibitem[Chellal and Boughanem(2018)]{chellal2018optimization}
Abdelhamid Chellal and Mohand Boughanem.
\newblock Optimization framework model for retrospective tweet summarization.
\newblock In \emph{Proceedings of the 33rd Annual ACM Symposium on Applied
  Computing}, pages 704--711, 2018.

\bibitem[Choi et~al.(2020)Choi, Grover, Shu, and Ermon]{grover2019fair}
Kristy Choi, Aditya Grover, Rui Shu, and Stefano Ermon.
\newblock Fair generative modeling via weak supervision.
\newblock In \emph{ICML}, 2020.

\bibitem[Dash et~al.(2019)Dash, Shandilya, Biswas, Ghosh, Ghosh, and
  Chakraborty]{dash2019summarizing}
Abhisek Dash, Anurag Shandilya, Arindam Biswas, Kripabandhu Ghosh, Saptarshi
  Ghosh, and Abhijnan Chakraborty.
\newblock Summarizing user-generated textual content: Motivation and methods
  for fairness in algorithmic summaries.
\newblock \emph{Proceedings of the ACM on Human-Computer Interaction},
  3\penalty0 (CSCW):\penalty0 1--28, 2019.

\bibitem[Devlin et~al.(2019)Devlin, Chang, Lee, and Toutanova]{devlin2018bert}
Jacob Devlin, Ming-Wei Chang, Kenton Lee, and Kristina Toutanova.
\newblock Bert: Pre-training of deep bidirectional transformers for language
  understanding.
\newblock In \emph{Proceedings of the 2019 Conference of the North American
  Chapter of the Association for Computational Linguistics: Human Language
  Technologies}, 2019.

\bibitem[Doyle(2014)]{doyle2014mapping}
Gabriel Doyle.
\newblock Mapping dialectal variation by querying social media.
\newblock In \emph{Proceedings of the 14th Conference of the European Chapter
  of the Association for Computational Linguistics}, pages 98--106, 2014.

\bibitem[Eisenstein et~al.(2010)Eisenstein, O’Connor, Smith, and
  Xing]{eisenstein2010latent}
Jacob Eisenstein, Brendan O’Connor, Noah~A Smith, and Eric Xing.
\newblock A latent variable model for geographic lexical variation.
\newblock In \emph{Proceedings of the 2010 conference on empirical methods in
  natural language processing}, pages 1277--1287, 2010.

\bibitem[Erkan and Radev(2004)]{erkan2004lexrank}
G{\"u}nes Erkan and Dragomir~R Radev.
\newblock Lexrank: Graph-based lexical centrality as salience in text
  summarization.
\newblock \emph{Journal of artificial intelligence research}, 2004.

\bibitem[Godin(2019)]{godin2019}
Fr\'{e}deric Godin.
\newblock \emph{Improving and Interpreting Neural Networks for Word-Level
  Prediction Tasks in Natural Language Processing}.
\newblock PhD thesis, Ghent University, Belgium, 2019.

\bibitem[Goldstein and Carbonell(1998)]{carbinell2017use}
Jade Goldstein and Jaime Carbonell.
\newblock Summarization: Using mmr for diversity-based reranking and evaluating
  summaries.
\newblock Technical report, Carnegie-Mellon Univ PA Language Technology Inst,
  1998.

\bibitem[He and Duan(2018)]{he2018twitter}
Ruifang He and Xingyi Duan.
\newblock Twitter summarization based on social network and sparse
  reconstruction.
\newblock In \emph{Thirty-Second AAAI Conference on Artificial Intelligence},
  2018.

\bibitem[Hendricks et~al.(2018)Hendricks, Burns, Saenko, Darrell, and
  Rohrbach]{hendricks2018women}
Lisa~Anne Hendricks, Kaylee Burns, Kate Saenko, Trevor Darrell, and Anna
  Rohrbach.
\newblock Women also snowboard: Overcoming bias in captioning models.
\newblock In \emph{European Conference on Computer Vision}, pages 793--811.
  Springer, 2018.

\bibitem[Hermann et~al.(2015)Hermann, Kocisky, Grefenstette, Espeholt, Kay,
  Suleyman, and Blunsom]{hermann2015teaching}
Karl~Moritz Hermann, Tomas Kocisky, Edward Grefenstette, Lasse Espeholt, Will
  Kay, Mustafa Suleyman, and Phil Blunsom.
\newblock Teaching machines to read and comprehend.
\newblock In \emph{Advances in neural information processing systems}, pages
  1693--1701, 2015.

\bibitem[Huang et~al.(2016)Huang, Guo, Kasakoff, and
  Grieve]{huang2016understanding}
Yuan Huang, Diansheng Guo, Alice Kasakoff, and Jack Grieve.
\newblock Understanding us regional linguistic variation with twitter data
  analysis.
\newblock \emph{Computers, Environment and Urban Systems}, 59:\penalty0
  244--255, 2016.

\bibitem[Inouye and Kalita(2011)]{inouye2011comparing}
David Inouye and Jugal~K Kalita.
\newblock Comparing twitter summarization algorithms for multiple post
  summaries.
\newblock In \emph{2011 IEEE Third international conference on privacy,
  security, risk and trust and 2011 IEEE third international conference on
  social computing}, pages 298--306. IEEE, 2011.

\bibitem[Jadhav and Rajan(2018)]{jadhav2018extractive}
Aishwarya Jadhav and Vaibhav Rajan.
\newblock Extractive summarization with swap-net: Sentences and words from
  alternating pointer networks.
\newblock In \emph{Proceedings of the 56th Annual Meeting of the Association
  for Computational Linguistics}, 2018.

\bibitem[Jones et~al.(2019)Jones, Kalbfeld, Hancock, and
  Clark]{jones2019testifying}
Taylor Jones, Jessica~Rose Kalbfeld, Ryan Hancock, and Robin Clark.
\newblock Testifying while black: An experimental study of court reporter
  accuracy in transcription of african american english.
\newblock \emph{Language}, 95\penalty0 (2):\penalty0 e216--e252, 2019.

\bibitem[J{\o}rgensen et~al.(2015)J{\o}rgensen, Hovy, and
  S{\o}gaard]{jorgensen2015challenges}
Anna J{\o}rgensen, Dirk Hovy, and Anders S{\o}gaard.
\newblock Challenges of studying and processing dialects in social media.
\newblock In \emph{Proceedings of the Workshop on Noisy User-generated Text},
  pages 9--18, 2015.

\bibitem[Kay et~al.(2015)Kay, Matuszek, and Munson]{kay2015unequal}
Matthew Kay, Cynthia Matuszek, and Sean~A Munson.
\newblock Unequal representation and gender stereotypes in image search results
  for occupations.
\newblock In \emph{Proceedings of the ACM Conference on Human Factors in
  Computing Systems}, 2015.

\bibitem[Kiritchenko and Mohammad(2018)]{kiritchenko2018examining}
Svetlana Kiritchenko and Saif~M Mohammad.
\newblock Examining gender and race bias in two hundred sentiment analysis
  systems.
\newblock \emph{NAACL HLT 2018}, 2018.

\bibitem[Kryscinski et~al.(2019)Kryscinski, Keskar, McCann, Xiong, and
  Socher]{kryscinski2019neural}
Wojciech Kryscinski, Nitish~Shirish Keskar, Bryan McCann, Caiming Xiong, and
  Richard Socher.
\newblock Neural text summarization: A critical evaluation.
\newblock In \emph{Proceedings of the 2019 Conference on Empirical Methods in
  Natural Language Processing and the 9th International Joint Conference on
  Natural Language Processing}, 2019.

\bibitem[Kulesza and Taskar(2012)]{kulesza2012determinantal}
Alex Kulesza and Ben Taskar.
\newblock Determinantal point processes for machine learning.
\newblock \emph{arXiv preprint arXiv:1207.6083}, 2012.

\bibitem[Lavan(2015)]{lavan2015negro}
Makeba Lavan.
\newblock The negro tweets his presence: Black twitter as social and political
  watchdog.
\newblock \emph{Modern Language Studies}, pages 56--65, 2015.

\bibitem[Lee et~al.(2009)Lee, Park, Ahn, and Kim]{lee2009automatic}
Ju-Hong Lee, Sun Park, Chan-Min Ahn, and Daeho Kim.
\newblock Automatic generic document summarization based on non-negative matrix
  factorization.
\newblock \emph{Information Processing \& Management}, 45\penalty0
  (1):\penalty0 20--34, 2009.

\bibitem[Lin and Hovy(2003)]{lin2003automatic}
Chin-Yew Lin and Eduard Hovy.
\newblock Automatic evaluation of summaries using n-gram co-occurrence
  statistics.
\newblock In \emph{Proceedings of the 2003 Human Language Technology Conference
  of the North American Chapter of the Association for Computational
  Linguistics}, pages 150--157, 2003.

\bibitem[Lin and Bilmes(2011)]{lin2011class}
Hui Lin and Jeff Bilmes.
\newblock A class of submodular functions for document summarization.
\newblock In \emph{Proceedings of the 49th Annual Meeting of the Association
  for Computational Linguistics: Human Language Technologies-Volume 1}, pages
  510--520. Association for Computational Linguistics, 2011.

\bibitem[Lin and Ng(2019)]{lin2019abstractive}
Hui Lin and Vincent Ng.
\newblock Abstractive summarization: A survey of the state of the art.
\newblock In \emph{Proceedings of the AAAI Conference on Artificial
  Intelligence}, 2019.

\bibitem[Liu and Lapata(2019)]{liu2019text}
Yang Liu and Mirella Lapata.
\newblock Text summarization with pretrained encoders.
\newblock In \emph{Proceedings of Conference on Empirical Methods in Natural
  Language Processing and International Joint Conference on Natural Language
  Processing}, 2019.

\bibitem[Lu et~al.(2018)Lu, Mardziel, Wu, Amancharla, and Datta]{lu2018gender}
Kaiji Lu, Piotr Mardziel, Fangjing Wu, Preetam Amancharla, and Anupam Datta.
\newblock Gender bias in neural natural language processing.
\newblock \emph{arXiv preprint arXiv:1807.11714}, 2018.

\bibitem[Luhn(1957)]{luhn1957statistical}
Hans~Peter Luhn.
\newblock A statistical approach to mechanized encoding and searching of
  literary information.
\newblock In \emph{IBM Journal of research and development}, 1957.

\bibitem[May et~al.(2019)May, Wang, Bordia, Bowman, and
  Rudinger]{may2019measuring}
Chandler May, Alex Wang, Shikha Bordia, Samuel Bowman, and Rachel Rudinger.
\newblock On measuring social biases in sentence encoders.
\newblock In \emph{Proceedings of the Conference of the North American Chapter
  of the Association for Computational Linguistics: Human Language
  Technologies}, 2019.

\bibitem[Mihalcea and Tarau(2004)]{mihalcea2004textrank}
Rada Mihalcea and Paul Tarau.
\newblock Textrank: Bringing order into text.
\newblock In \emph{Proceedings of the conference on empirical methods in
  natural language processing}, 2004.

\bibitem[Mikolov et~al.(2015)Mikolov, Chen, Corrado, and
  Dean]{mikolov2015computing}
Tomas Mikolov, Kai Chen, Gregory~S Corrado, and Jeffrey~A Dean.
\newblock Computing numeric representations of words in a high-dimensional
  space, May~19 2015.
\newblock US Patent 9,037,464.

\bibitem[Miller(2019)]{miller2019leveraging}
Derek Miller.
\newblock Leveraging bert for extractive text summarization on lectures.
\newblock \emph{arXiv preprint arXiv:1906.04165}, 2019.

\bibitem[Miller et~al.(2012)Miller, Dickinson, and Hu]{miller2012gender}
Zachary Miller, Brian Dickinson, and Wei Hu.
\newblock Gender prediction on twitter using stream algorithms with n-gram
  character features.
\newblock 2012.

\bibitem[Nadeem et~al.(2020)Nadeem, Bethke, and Reddy]{nadeem2020stereoset}
Moin Nadeem, Anna Bethke, and Siva Reddy.
\newblock Stereoset: Measuring stereotypical bias in pretrained language
  models, 2020.

\bibitem[Nallapati et~al.(2017)Nallapati, Zhai, and
  Zhou]{nallapati2017summarunner}
Ramesh Nallapati, Feifei Zhai, and Bowen Zhou.
\newblock Summarunner: A recurrent neural network based sequence model for
  extractive summarization of documents.
\newblock In \emph{Thirty-First AAAI Conference on Artificial Intelligence},
  2017.

\bibitem[Nguyen et~al.(2018)Nguyen, Lai, Nguyen, and Le~Nguyen]{nguyen2018tsix}
Minh-Tien Nguyen, Dac~Viet Lai, Huy~Tien Nguyen, and Minh Le~Nguyen.
\newblock Tsix: a human-involved-creation dataset for tweet summarization.
\newblock In \emph{Proceedings of the Eleventh International Conference on
  Language Resources and Evaluation}, 2018.

\bibitem[Ott(2016)]{ott2016tweet}
Margaret Ott.
\newblock Tweet like a girl: A corpus analysis of gendered language in social
  media.
\newblock \emph{Yale University, apr}, 2016.

\bibitem[Ozsoy et~al.(2011)Ozsoy, Alpaslan, and Cicekli]{ozsoy2011text}
Makbule~Gulcin Ozsoy, Ferda~Nur Alpaslan, and Ilyas Cicekli.
\newblock Text summarization using latent semantic analysis.
\newblock \emph{Journal of Information Science}, 2011.

\bibitem[Padmakumar and Saran(2016)]{padmakumar2016unsupervised}
Aishwarya Padmakumar and Akanksha Saran.
\newblock Unsupervised text summarization using sentence embeddings.
\newblock Technical report, Technical Report, University of Texas at Austin,
  2016.

\bibitem[Park et~al.(2018)Park, Shin, and Fung]{park2018reducing}
Ji~Ho Park, Jamin Shin, and Pascale Fung.
\newblock Reducing gender bias in abusive language detection.
\newblock In \emph{Proceedings of EMNLP 2018}, 2018.

\bibitem[Radev et~al.(2001)Radev, Blair-Goldensohn, and
  Zhang]{radev2001experiments}
Dragomir~R Radev, Sasha Blair-Goldensohn, and Zhu Zhang.
\newblock Experiments in single and multidocument summarization using mead.
\newblock In \emph{First document understanding conference}, page 1{\`A}8.
  Citeseer, 2001.

\bibitem[Rickford(2016)]{rickford2016raciolinguistics}
John~R Rickford.
\newblock \emph{Raciolinguistics: How language shapes our ideas about race}.
\newblock Oxford University Press, 2016.

\bibitem[Rosenberg and Hirschberg(2007)]{rosenberg2007v}
Andrew Rosenberg and Julia Hirschberg.
\newblock V-measure: A conditional entropy-based external cluster evaluation
  measure.
\newblock In \emph{Proceedings of the 2007 joint conference on empirical
  methods in natural language processing and computational natural language
  learning (EMNLP-CoNLL)}, pages 410--420, 2007.

\bibitem[Rossiello et~al.(2017)Rossiello, Basile, and
  Semeraro]{rossiello2017centroid}
Gaetano Rossiello, Pierpaolo Basile, and Giovanni Semeraro.
\newblock Centroid-based text summarization through compositionality of word
  embeddings.
\newblock In \emph{Proceedings of the MultiLing 2017 Workshop on Summarization
  and Summary Evaluation Across Source Types and Genres}, pages 12--21, 2017.

\bibitem[Sap et~al.(2019)Sap, Card, Gabriel, Choi, and Smith]{sap2019risk}
Maarten Sap, Dallas Card, Saadia Gabriel, Yejin Choi, and Noah~A Smith.
\newblock The risk of racial bias in hate speech detection.
\newblock In \emph{Proceedings of the 57th Annual Meeting of the Association
  for Computational Linguistics}, pages 1668--1678, 2019.

\bibitem[Sassaman et~al.(2020)Sassaman, Lee, Irvine, and
  Narayan]{sassaman2020creating}
Hannah Sassaman, Jennifer Lee, Jenessa Irvine, and Shankar Narayan.
\newblock Creating community-based tech policy: case studies, lessons learned,
  and what technologists and communities can do together.
\newblock In \emph{Proceedings of the 2020 Conference on Fairness,
  Accountability, and Transparency}, pages 685--685, 2020.

\bibitem[Snyder et~al.(1977)Snyder, Tanke, and Berscheid]{snyder1977social}
Mark Snyder, Elizabeth~Decker Tanke, and Ellen Berscheid.
\newblock Social perception and interpersonal behavior: On the self-fulfilling
  nature of social stereotypes.
\newblock \emph{Journal of Personality and social Psychology}, 35\penalty0
  (9):\penalty0 656, 1977.

\bibitem[Sun et~al.(2019)Sun, Gaut, Tang, Huang, ElSherief, Zhao, Mirza,
  Belding, Chang, and Wang]{sun2019mitigating}
Tony Sun, Andrew Gaut, Shirlyn Tang, Yuxin Huang, Mai ElSherief, Jieyu Zhao,
  Diba Mirza, Elizabeth Belding, Kai-Wei Chang, and William~Yang Wang.
\newblock Mitigating gender bias in natural language processing: Literature
  review.
\newblock \emph{arXiv preprint arXiv:1906.08976}, 2019.

\bibitem[Tan and Celis(2019)]{tan2019assessing}
Yi~Chern Tan and L~Elisa Celis.
\newblock Assessing social and intersectional biases in contextualized word
  representations.
\newblock In \emph{Neural Information Processing Systems}, 2019.

\bibitem[Tatman(2017)]{tatman2017gender}
Rachael Tatman.
\newblock Gender and dialect bias in youtube’s automatic captions.
\newblock In \emph{Proceedings of the First ACL Workshop on Ethics in Natural
  Language Processing}, 2017.

\bibitem[Terry et~al.(2010)Terry, Hendrick, Evangelou, and
  Smith]{terry2010variable}
J~Michael Terry, Randall Hendrick, Evangelos Evangelou, and Richard~L Smith.
\newblock Variable dialect switching among african american children:
  Inferences about working memory.
\newblock \emph{Lingua}, 120\penalty0 (10):\penalty0 2463--2475, 2010.

\bibitem[Vanderwende et~al.(2007)Vanderwende, Suzuki, Brockett, and
  Nenkova]{vanderwende2007beyond}
Lucy Vanderwende, Hisami Suzuki, Chris Brockett, and Ani Nenkova.
\newblock Beyond sumbasic: Task-focused summarization with sentence
  simplification and lexical expansion.
\newblock \emph{Information Processing \& Management}, 43\penalty0
  (6):\penalty0 1606--1618, 2007.

\bibitem[Zhao et~al.(2017)Zhao, Wang, Yatskar, Ordonez, and Chang]{zhao2017men}
Jieyu Zhao, Tianlu Wang, Mark Yatskar, Vicente Ordonez, and Kai-Wei Chang.
\newblock Men also like shopping: Reducing gender bias amplification using
  corpus-level constraints.
\newblock In \emph{Proceedings of the 2017 Conference on Empirical Methods in
  Natural Language Processing}, 2017.

\bibitem[Zhao et~al.(2018)Zhao, Zhou, Li, Wang, and Chang]{zhao2018learning}
Jieyu Zhao, Yichao Zhou, Zeyu Li, Wei Wang, and Kai-Wei~Chang Chang.
\newblock Learning gender-neutral word embeddings.
\newblock In \emph{Proceedings of the 2018 Conference on Empirical Methods in
  Natural Language Processing}, 2018.

\bibitem[Zhong et~al.(2019)Zhong, Liu, Wang, Qiu, and
  Huang]{zhong2019searching}
Ming Zhong, Pengfei Liu, Danqing Wang, Xipeng Qiu, and Xuan-Jing Huang.
\newblock Searching for effective neural extractive summarization: What works
  and what’s next.
\newblock In \emph{Proceedings of the 57th Annual Meeting of the Association
  for Computational Linguistics}, pages 1049--1058, 2019.

\bibitem[Zhong et~al.(2020)Zhong, Liu, Chen, Wang, Qiu, and
  Huang]{zhong2020extractive}
Ming Zhong, Pengfei Liu, Yiran Chen, Danqing Wang, Xipeng Qiu, and Xuan-Jing
  Huang.
\newblock Extractive summarization as text matching.
\newblock In \emph{Proceedings of the 58th Annual Meeting of the Association
  for Computational Linguistics}, pages 6197--6208, 2020.

\end{thebibliography}

\appendix

\clearpage

\section{{Details of summarization algorithms}} \label{sec:algo_details}

\noindent
\textbf{TF-IDF:} This baseline \cite{luhn1957statistical} uses frequency of the words in a sentence to quantify their weight.
However, if a word is very common and occurs in a lot of sentences, then it is likely that the word is part of the grammar structure; hence inverse of document frequency is also taken into account while calculating its score\footnote{\label{ftn:impl} Internally implemented using the python sklearn and networkx libraries.}.
For any sentence $x$ in $S$, let $W(x)$ denote the set of words in the sentence. Then the weight assigned to $x$ is
$ \frac{1}{|W(x)|}\sum_{w \in W(x)} tf(w, x) \cdot \log \frac{|S|}{idf(w, S)},$
where $tf(w,x)$ is the number of times $w$ occurs in $x$ and $idf(w, S)$ is the number of sentences in which $w$ occurs.

\begin{figure*}[t]
\includegraphics[width=\linewidth]{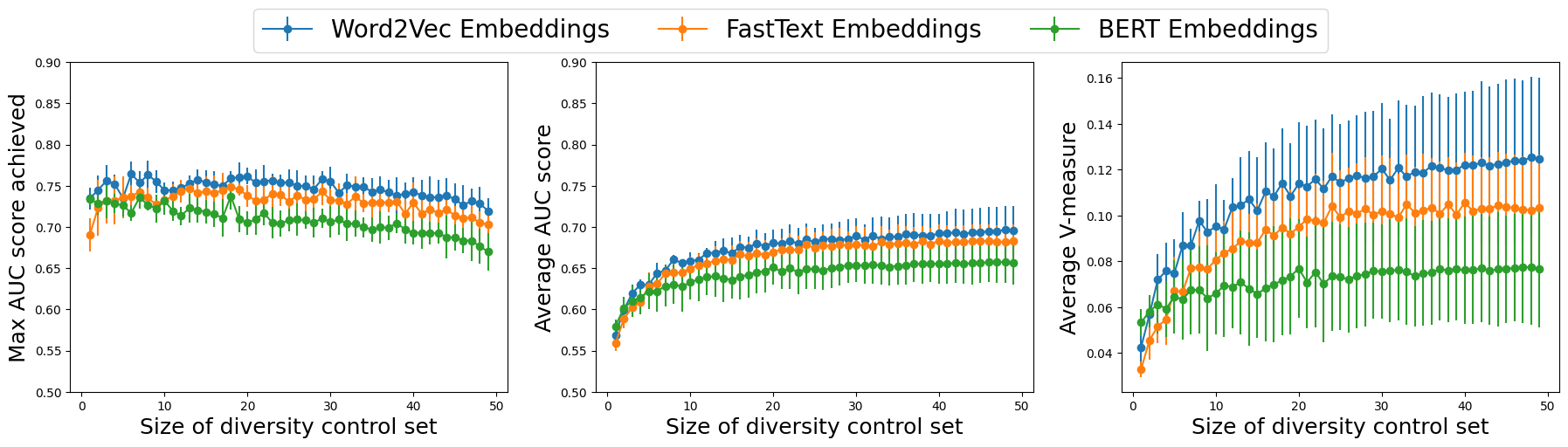}
  \subfloat[Maximum AUC score vs $|T|$]{\hspace{.33\linewidth}}
\subfloat[Mean AUC score score vs $|T|$]{\hspace{.33\linewidth}}
  \subfloat[Maximum V-measure score vs $|T|$]{\hspace{.33\linewidth}}
\caption{\small
The figure presents how effective different diversity control sets are in clustering posts of the different dialects. 
Figure (a) presents the average maximum AUC score achieved by a control set across folds for different control set sizes, while Figure (b) presents the mean AUC score achieved by a control set across folds.
As an alternative measure, Figure (c) presents the mean V-measure across folds.
}
\label{fig:control_set_analysis}
\end{figure*}

\noindent
\textbf{Hybrid TF-IDF:} The standard TF-IDF has been noted to have poor performance for Twitter posts, primarily due to lack of generalization of Twitter posts as documents \cite{nguyen2018tsix}. 
Correspondingly, a Hhbrid TF-IDF \cite{inouye2011comparing} approach is proposed that
calculates word frequency considering the entire collection as a single document.\cref{ftn:impl}
In other words, the $tf(w, x)$ term in the weight assigned by TF-IDF is replaced by $tf(w, S)$ for Hybrid TF-IDF.

\noindent
\textbf{LexRank:} This unsupervised summarizer constructs a graph over the dataset, with similarity between sentences quantifying the edge-weights \cite{erkan2004lexrank}, 
%
measured using cosine distance between their TF-IDF word vectors.
Using the PageRank algorithm, sentences are then ranked based on how ``central'' they are within the graph \footnote{\url{https://github.com/crabcamp/lexrank}}. 

\noindent
\textbf{TextRank:} This algorithm quantifies the similarity using a modified score of word document frequency \cite{mihalcea2004textrank} and then uses PageRank to rank the sentences; however, it has been shown to achieve better performance for some standard datasets \cite{nguyen2018tsix}\cref{ftn:impl}.

\noindent
\textbf{Centroid-Word2Vec}: This algorithm 
assigns importance scores to sentences based on their distance from the \textit{centroid} of the dataset \cite{rossiello2017centroid} (related to \cite{miller2019leveraging}).
For vector representation, we use Word2Vec embeddings, pre-trained on a large Twitter dataset \cite{godin2019}.
As mentioned in Section~\ref{sec:introduction}, it also has a non-redundancy component; %
if the minimum distance between the feature of candidate sentence and the feature of a sentence already in the summary is higher than a threshold (0.95 in our case), it is discarded\footnote{\url{https://github.com/TextSummarizer/TextSummarizer}}.

\noindent
\textbf{MMR:} This is a post-processing re-ranking algorithm that, at every iteration, greedily chooses the sentence which has highest MMR score, calculated as the combination of importance score and dissimilarity with the sentences already present in the summary \cite{carbinell2017use, lin2011class}. 
To get the base importance score, we use the TF-IDF algorithm \cref{ftn:impl}.
Since MMR is a greedy post-processing approach itself, we do not use it as a blackbox algorithm for our framework.

\noindent
\textbf{SummaRuNNer: } Finally, we use a recent Recurrent Neural Network based method, SummaRuNNer \cite{nallapati2017summarunner}, that treats summarization as a sequential classification problem over the dataset, and generates summaries comparable to state-of-the-art for the CNN/DailyMail dataset \cite{hermann2015teaching}.
Since it is not possible to train this model over the Twitter datasets we consider (due to non-availability of dataset-summary pairs for Twitter dataset), we use the model pre-trained on a standard summarization evaluation dataset \footnote{Unofficial implementation: \url{https://github.com/hpzhao/SummaRuNNer}}.

Inouye and Kalita \cite{inouye2011comparing} empirically analyze the performance of TF-IDF, Hybrid TF-IDF, LexRank and TextRank on small Twitter datasets (containing only around 1500 tweets for 50 trending topics, not sufficient for a diversity analysis).
Their findings suggest that 
Hybrid TF-IDF produce better summaries for Twitter summarization than TF-IDF, LexRank and TextRank (as evaluated using ROUGE metrics and manually-generated summaries).
For larger and more-recent Twitter datasets, Nguyen et al. \cite{nguyen2018tsix} found that TextRank and Hybrid TF-IDF have similar performance.
Rossiello et al. \cite{rossiello2017centroid} showed that the centroid-based approach performs better than LexRank, frequency and RNN-based models on the DUC-2004 dataset.
The original papers for most of these algorithms primarily focussed on evaluation of these methods on DUC tasks or CNN/DailyMail datasets; however, the documents in these datasets correspond to news articles from a particular agency and do not usually have significant dialect diversity within them.

 \begin{table}[t]
\footnotesize
  \caption{Diversity control set for \textit{TwitterAAE evaluations}} 
\begin{tabular}{|p{\linewidth}|}
    \hline
     \textbf{AAE tweets}  \\
     \hline 
    ``ATMENTION yea dats more like it b4 I make a trip up der'' \\
  ``these n***s talmbout money but . really ain't getting no money .. I be laughing at these n***s cause that shit funny ATMENTION'' \\
  ``Me and Pay got matching coupes, me and kid fucked ya boo'' \\
  ``ATMENTION he bites his lips and manages to kick off his remaining clothes'' \\
  ``Our Dog Is A Big Baby And A Wanna Be Thug EMOJI'' \\
  ``Its a Damn Shame' iont GangBang but i beat a N*** Blue Black'' \\
  ``ATMENTION yes, my amazon . Lol Im good . Pop-a-lock came by . Thx!'' \\
  ``ATMENTION: ATMENTION You talking now? RIGHT? im typing nd texting not talking'' \\
  ``Soon as u think you gotcha 1 you find out she fckin erbody!!'' \\
  ``ATMENTION lmaooooooooooooooooooooooo, that was the funniest shit ever to hit twitter dawg :D swearrr .. But yall do yall thang'' \\
  ``Yea Ill Be Good In Bed But Ill Be Bad To Ya!'' \\
  ``ATMENTION nope tell her get dressed im bouta come get her lol'' \\
  ``Now omw to get my hair done for coronation tomorrow'' \\
  ``Ohhhh Hell Naw Dis B**** Shay Got My Last Name * Johnson *'' \\
    \hline 
    \hline 
     \textbf{WHE tweets}  \\
     \hline 
  ``You don't have to keep on smiling that smile that's driving me wild'' \\
  ``ATMENTION it's probably dead because he hasn't texted me back either'' \\
  ``ATMENTION amen . Honestly have trouble watching that movie . Just because of her.'' \\
  ``I need to get on a laptop so I can change my tumblr bio'' \\
  ``Shout out to the blue collar workers . Gotta love it'' \\
  ``Jax keeps curling up on my bed and tossing and turning repeatedly . Like he cant get comfy . \#Soocute \#Puppylove'' \\
  ``ATMENTION you just can't go wrong with Chili's . They serve a mean chips and salsa'' \\
  ``ATMENTION Tenuta hasn't been good since he left GT and he hates recruiting'' \\
  ``ATMENTION: Probably the coolest thing I can do ATMENTION yeah, pretty frickin' sweet! Thanks'' \\
  ``ATMENTION you said we were hanging all day...Lol I don't have a car alslo'' \\
  ``I want a love like off The Vow .. \#perfect \#oneday'' \\
  ``Philosophy is the worst thing to ever happen to the world'' \\
  ``How come I can never get in a " gunning " fight with anyone? \#Jealous'' \\
  ``'Poor poor Merle, bravo for Michael Rooker and Norman Reedus's performance on last night's show.' \\
  \hline
  \end{tabular}
  \label{tbl:aae_div_set}
\end{table}




\section{Choice of diversity control set} \label{sec:div_control_analysis}
In this section, we provide a method to construct a \textit{good} diversity control set.
For this analysis, we limit ourselves to assessing diversity with respect to AAE and WHE dialects.
We employ a smaller processed version of TwitterAAE dataset, containing 250 AAE posts and 250 WHE (provided by \cite{blodgett2018twitter}),
to select diversity control sets.

\noindent
\paragraph{Evaluation details.}
The size of the diversity control set should ideally be much smaller than the evaluation dataset; this will assist in better curation of the control sets.
Hence, we restrict the size of the control sets for our simulations to be atmost 50.

We perform a 5-fold cross validation setup for this simulation.
For each fold, we have a validation partition $U$ of 400 posts and a train partition of 100 posts (both containing equal number of AAE and WHE posts); we use the train partition to construct a diversity control set.
We sample a set of posts from the train partition, making sure that the set has equal number of AAE and WHE posts, and use it as diversity control set; let $T$ denote this set of posts.
Then for each $z \in T$ and $x \in U$, we calculate the score $ \sim(z, x)$, and to each $x \in U$, we assign the dialect label of the post $\arg \max_{z \in T}\sim(z, x)$.
Finally, for this prediction task, we report the AUC score and V-measure between the assigned and true dialect labels for posts in $U$. AUC refers to the area under the Receiver Operating Characteristic (ROC) curve. 
It is measure commonly used to evaluate how the performance of a binary learning task.
V-measure, on the other hand, is used to evaluate clustering tasks \cite{rosenberg2007v}. This measure combines homogeneity (the extent to which AAE clusters contain AAE posts) and completeness (all AAE posts are assigned to AAE clusters).
We repeat the sample-and-predict process 50 times for each fold, and we record the max, mean and standard error of AUC and V-measures across all repetitions.

To calculate similarity $\sim(z, x)$ between two sentences, we will use pre-trained word and sentence embeddings to find the feature vectors for these sentences, and then measure the similarity as $1 - \textrm{cosine-distance}$ between the feature vectors.
We employ three popular and robust pre-trained embeddings for this task: (a) Word2Vec \cite{mikolov2015computing}, (b) FastText \cite{bojanowski2017enriching}, and (c) BERT embeddings \cite{devlin2018bert}.
Using Word2Vec and FastText model, we obtain word representations; to obtain sentence embeddings from word representations, we use the aggregation method of Arora et al. \cite{arora2016simple} which computes the weighted average of the embeddings of the words in the sentence, where the weight assigned to a word is proportional to the smooth inverse frequency of the word.
For Word2Vec and FastText, we use the models pre-trained on a corpus of 400 million posts \cite{godin2019}.
Output from second-last hidden layer of the pre-trained BERT model can be used to directly obtain sentence embeddings.

\noindent
\paragraph{Results.}
Figure~\ref{fig:control_set_analysis} show that diversity control sets constructed in this manner are indeed suitable for differentiating between posts of different dialects.
Plot~\ref{fig:control_set_analysis}a shows that \textit{good control sets} are able to achieve AUC scores greater than 0.8 (including the one presented in Table~\ref{tbl:aae_div_set}).
Furthermore, the average AUC score is also greater than 0.65 for diversity control set sizes greater than 10, implying that diversity control sets of size between 10 and 50 are indeed suitable for this task.
Given that the diversity control sets do perform fairly well on this clustering task, this provides further insight into the improved dialect diversity when using our post-processing framework with standard summarization algorithms as blackbox.
Secondly, Word2Vec embeddings achieve better performance than FastText and BERT embeddings and, hence, we use Word2vec representations for the empirical analysis of our framework as well.

Using the above method, we construct a diversity control set of size 28 for \textit{TwitterAAE Evaluations} (Table~\ref{tbl:aae_div_set}), a control set of size 40 for \textit{Crowdflower Evaluations} (Table~\ref{tbl:div_control_gender}), and control set of size 20 for \textit{Claritin Evaluations} (Table~\ref{tbl:div_control_claritin}).

\section{Other details and results for TwitterAAE dataset}
\label{sec:aae_other}
The control set used for TwitterAAE simulations is provided in Table~\ref{tbl:aae_div_set}.

\subsection{Evaluation of our model on random collections of TwitterAAE datasets} \label{sec:aae_all_results}
\begin{figure*}[!htbp]
\small
\includegraphics[width=\linewidth]{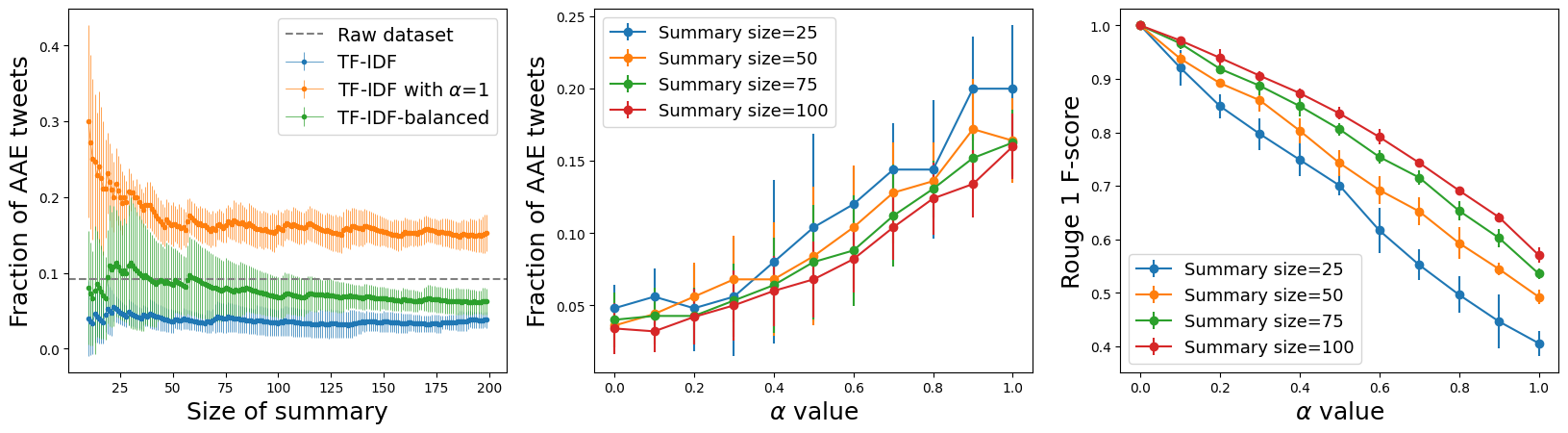}
  \subfloat[AAE fraction vs summary size]{\hspace{.33\linewidth}}
\subfloat[AAE fraction in summary vs $\alpha$]{\hspace{.33\linewidth}}
  \subfloat[Rouge-1 F-score vs $\alpha$]{\hspace{.33\linewidth}}
\caption{Evaluation of our model on datasets containing 8.7\% AAE tweets using TF-IDF as algorithm $A$.
}
\label{fig:random_aae_frac_fair_09_tfidf}
\end{figure*}

\begin{figure*}[!htbp]
\small
\includegraphics[width=\linewidth]{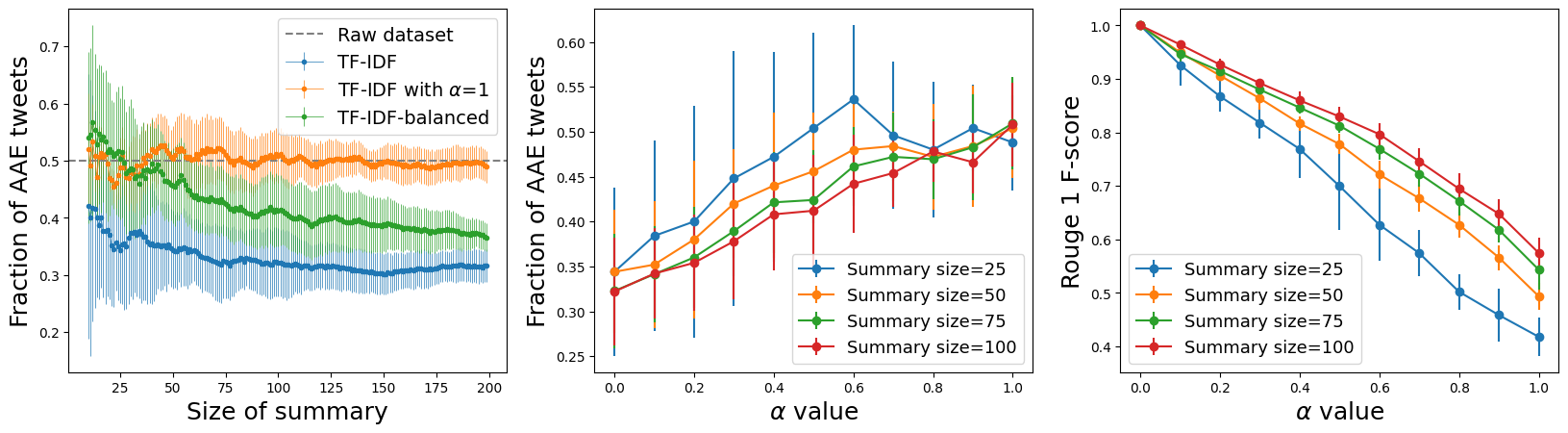}
  \subfloat[AAE fraction vs summary size]{\hspace{.33\linewidth}}
\subfloat[AAE fraction in summary vs $\alpha$]{\hspace{.33\linewidth}}
  \subfloat[Rouge-1 F-score vs $\alpha$]{\hspace{.33\linewidth}}
\caption{Evaluation of our model on datasets containing 50\% AAE tweets using TF-IDF as algorithm $A$.
}
\label{fig:random_aae_frac_fair_50_tfidf}
\end{figure*}

\begin{figure*}[!htbp]
\small
\includegraphics[width=\linewidth]{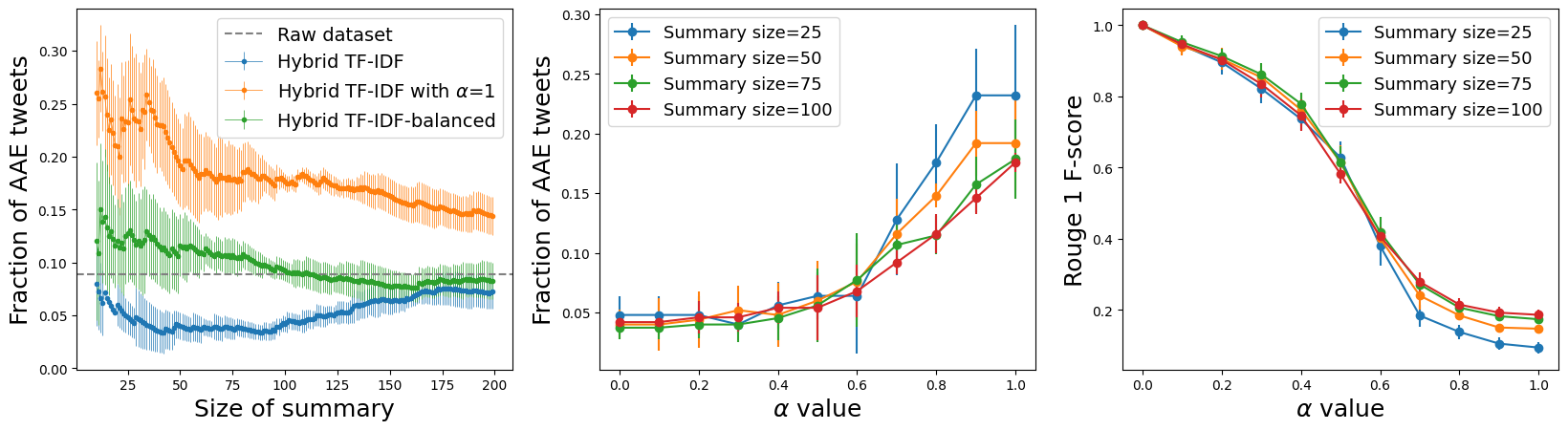}
  \subfloat[AAE fraction vs summary size]{\hspace{.33\linewidth}}
\subfloat[AAE fraction in summary vs $\alpha$]{\hspace{.33\linewidth}}
  \subfloat[Rouge-1 F-score vs $\alpha$]{\hspace{.33\linewidth}}
\caption{Evaluation of our model on datasets containing 8.7\% AAE tweets using Hybrid TF-IDF as algorithm $A$. Here $\alpha=0.7$ for balanced algorithm
}
\label{fig:random_aae_frac_fair_09_hybtfidf}
\end{figure*}

\begin{figure*}[!htbp]
\small
\includegraphics[width=\linewidth]{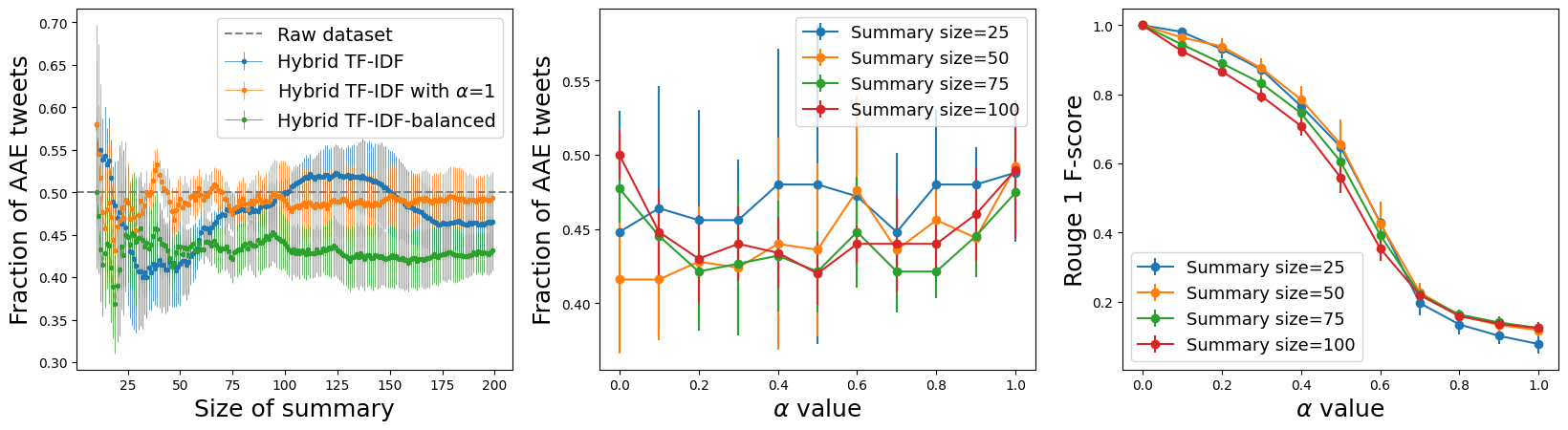}
  \subfloat[AAE fraction  vs summary size]{\hspace{.33\linewidth}}
\subfloat[AAE fraction in summary vs $\alpha$]{\hspace{.33\linewidth}}
  \subfloat[Rouge-1 F-score vs $\alpha$]{\hspace{.33\linewidth}}
\caption{Evaluation of our model on datasets containing 50\% AAE tweets using Hybrid TF-IDF as algorithm $A$. Here $\alpha=0.7$ for balanced algorithm.
}
\label{fig:random_aae_frac_fair_50_hybtfidf}
\end{figure*}

\begin{figure*}[!htbp]
\small
\includegraphics[width=\linewidth]{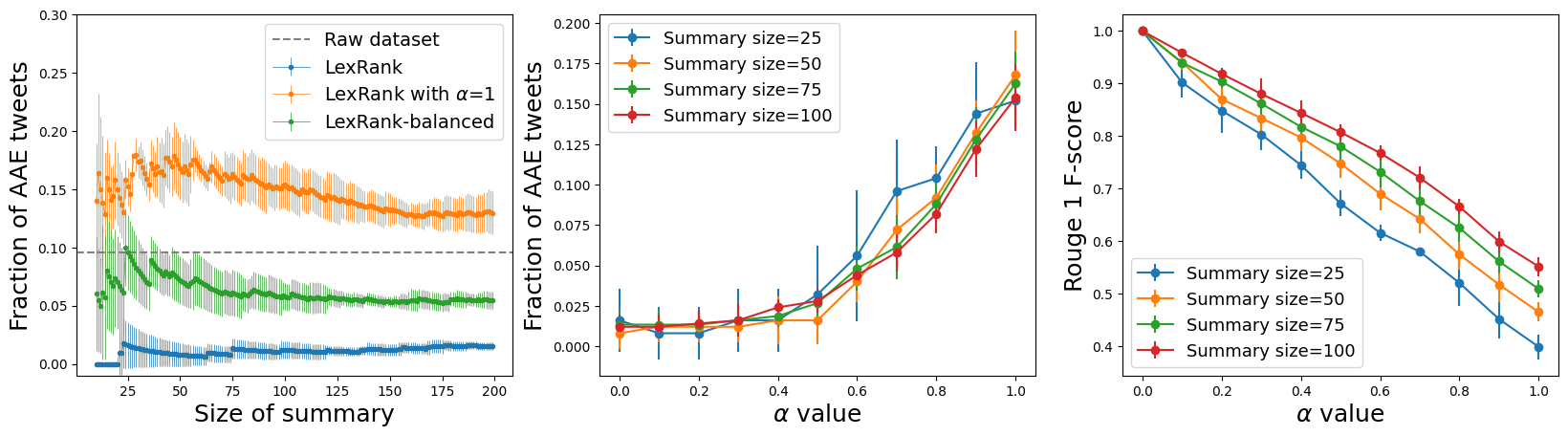}
  \subfloat[AAE fraction vs summary size]{\hspace{.33\linewidth}}
\subfloat[AAE fraction in summary vs $\alpha$]{\hspace{.33\linewidth}}
  \subfloat[Rouge-1 F-score vs $\alpha$]{\hspace{.33\linewidth}}
\caption{Evaluation of our model on datasets containing 8.7\% AAE tweets using LexRank as algorithm $A$. Here $\alpha=0.7$ for balanced algorithm.
}
\label{fig:random_aae_frac_fair_09_lexrank}
\end{figure*}

\begin{figure*}[!htbp]
\small
\includegraphics[width=\linewidth]{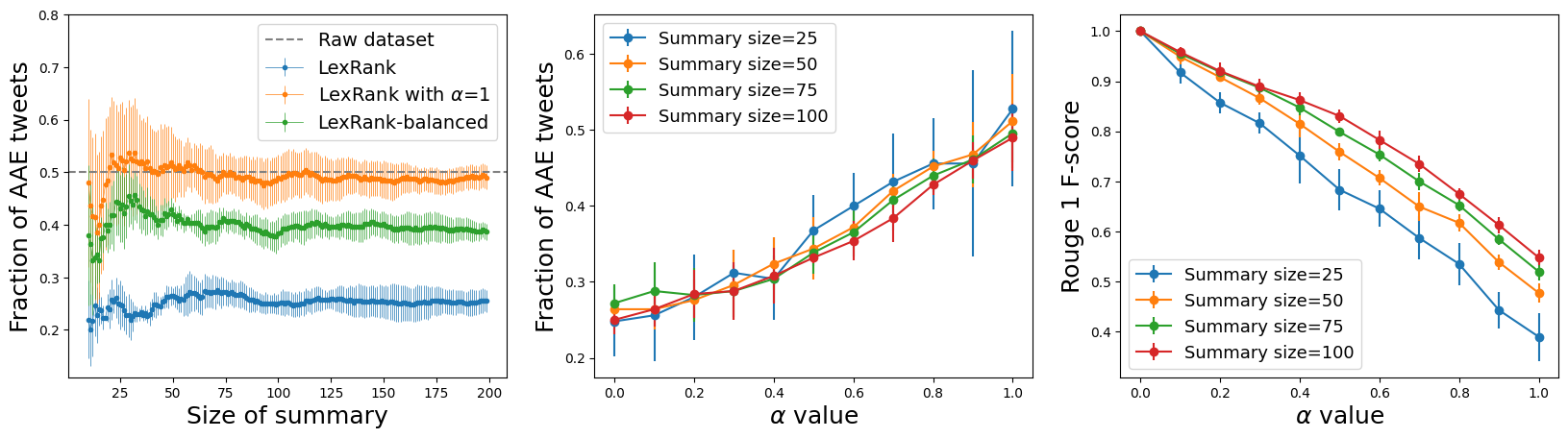}
  \subfloat[AAE fraction  vs summary size]{\hspace{.33\linewidth}}
\subfloat[AAE fraction in summary vs $\alpha$]{\hspace{.33\linewidth}}
  \subfloat[Rouge-1 F-score vs $\alpha$]{\hspace{.33\linewidth}}
\caption{Evaluation of our model on datasets containing 50\% AAE tweets using LexRank as algorithm $A$. Here $\alpha=0.7$ for balanced algorithm.
}
\label{fig:random_aae_frac_fair_50_lexrank}
\end{figure*}

\begin{figure*}[!htbp]
\small
\includegraphics[width=\linewidth]{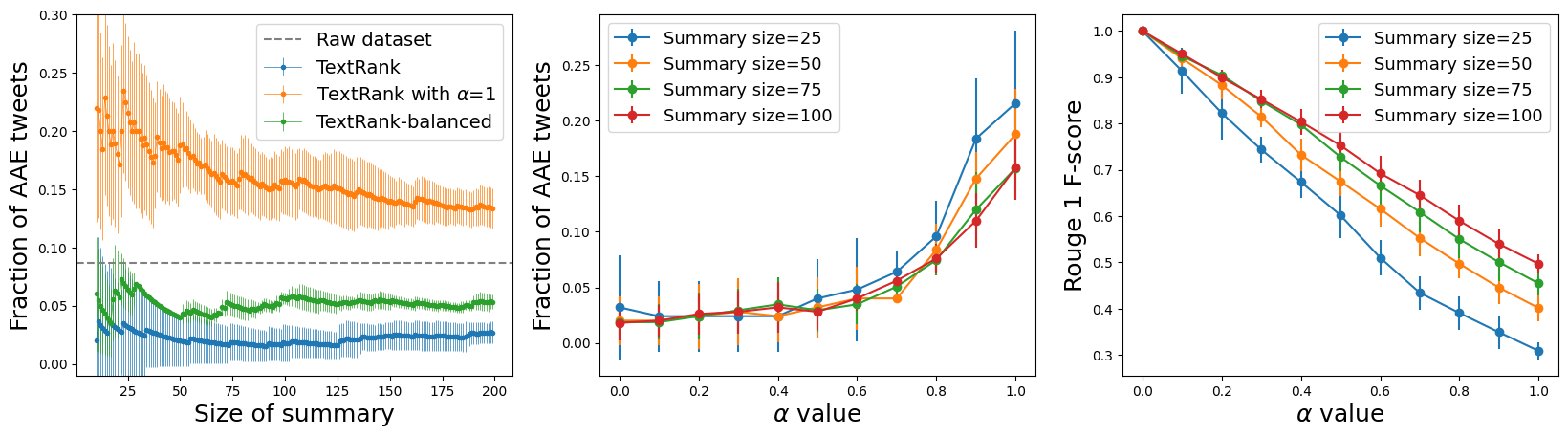}
  \subfloat[AAE fraction vs summary size]{\hspace{.33\linewidth}}
\subfloat[AAE fraction in summary vs $\alpha$]{\hspace{.33\linewidth}}
  \subfloat[Rouge-1 F-score vs $\alpha$]{\hspace{.33\linewidth}}
\caption{Evaluation of our model on datasets containing 8.7\% AAE tweets using TextRank as algorithm $A$. Here $\alpha=0.7$ for balanced algorithm.
}
\label{fig:random_aae_frac_fair_09_textrank}
\end{figure*}

\begin{figure*}[!htbp]
\small
\includegraphics[width=\linewidth]{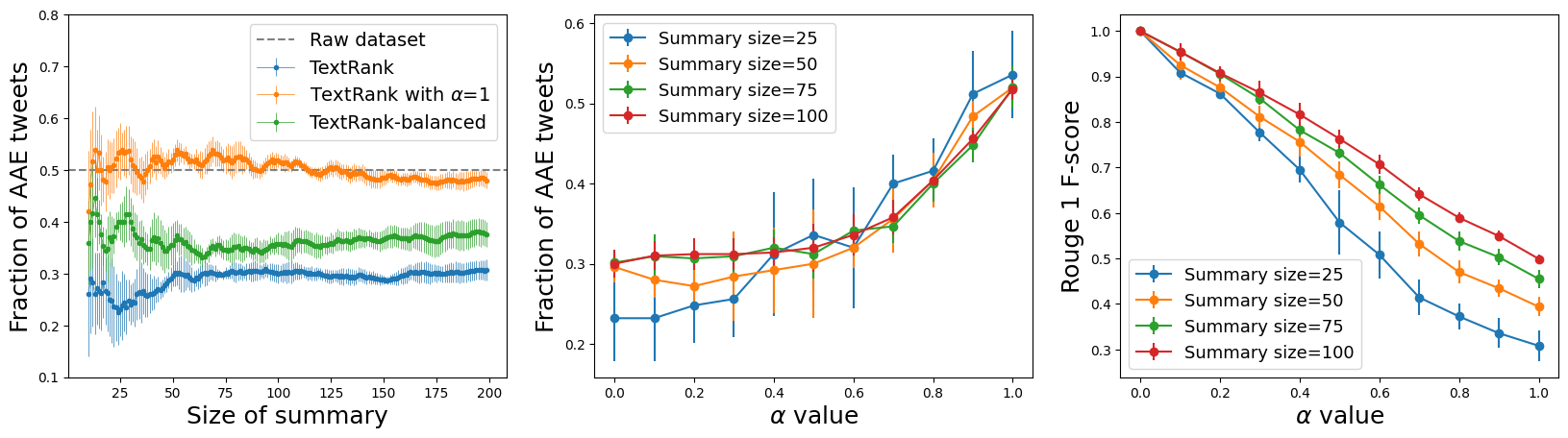}
  \subfloat[AAE fraction vs summary size]{\hspace{.33\linewidth}}
\subfloat[AAE fraction in summary vs $\alpha$]{\hspace{.33\linewidth}}
  \subfloat[Rouge-1 F-score vs $\alpha$]{\hspace{.33\linewidth}}
\caption{Evaluation of our model on datasets containing 50\% AAE tweets using TextRank as algorithm $A$. Here $\alpha=0.7$ for balanced algorithm.
}
\label{fig:random_aae_frac_fair_50_textrank}
\end{figure*}

\begin{figure*}[!htbp]
\small
\includegraphics[width=\linewidth]{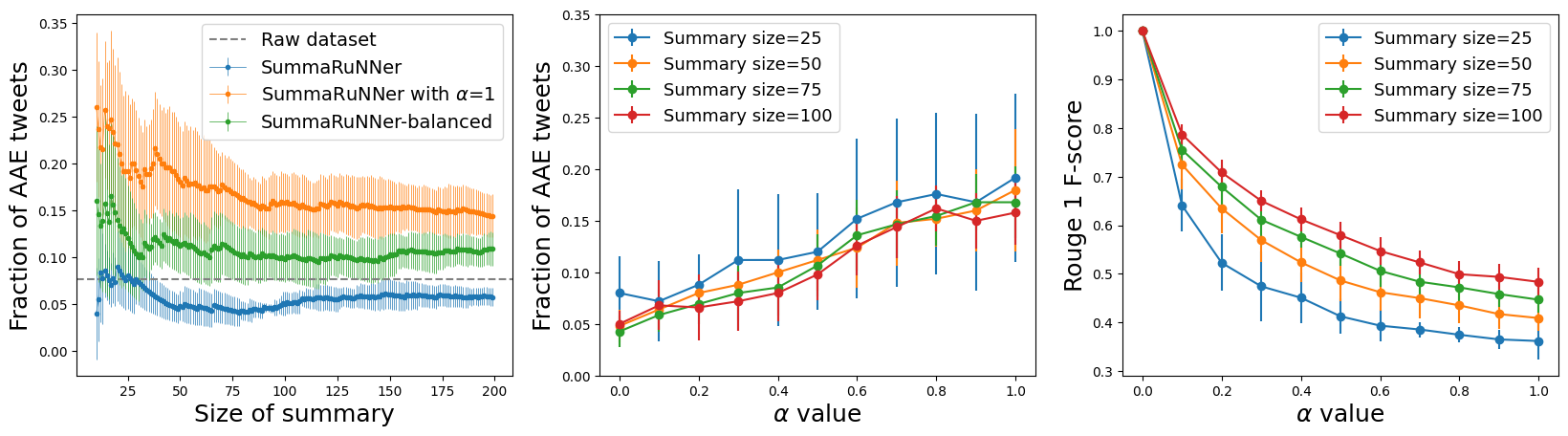}
  \subfloat[AAE fraction vs summary size]{\hspace{.33\linewidth}}
\subfloat[AAE fraction in summary vs $\alpha$]{\hspace{.33\linewidth}}
  \subfloat[Rouge-1 F-score vs $\alpha$]{\hspace{.33\linewidth}}
\caption{Evaluation of our model on datasets containing 8.7\% AAE tweets using SummaRuNNer as algorithm $A$.
}
\label{fig:random_aae_frac_fair_09_srnn}
\end{figure*}

\begin{figure*}[!htbp]
\small
\includegraphics[width=\linewidth]{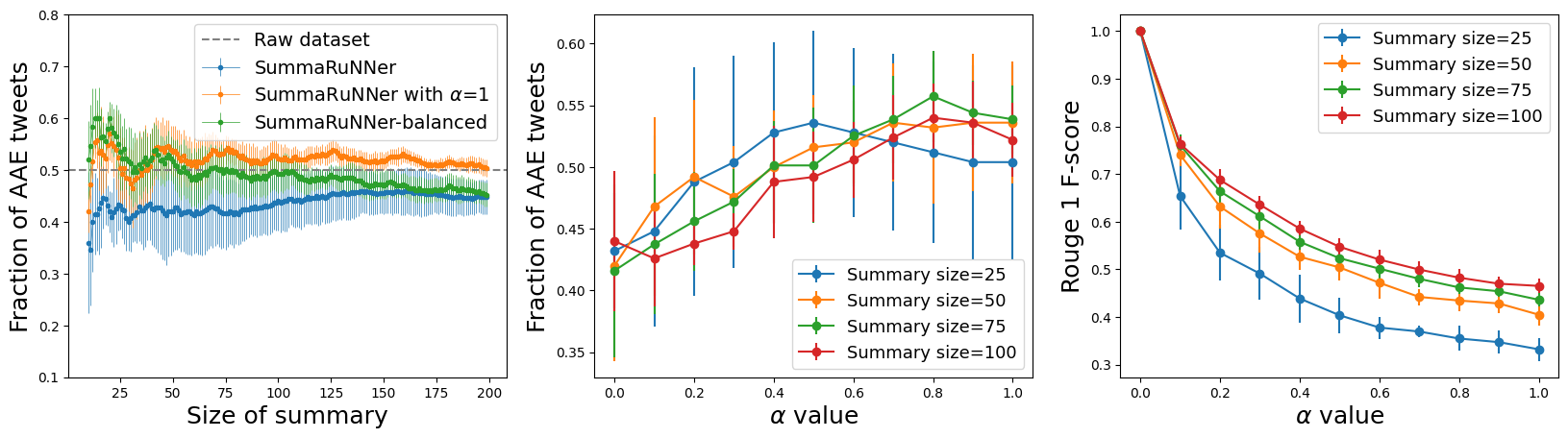}
  \subfloat[AAE fraction vs summary size]{\hspace{.33\linewidth}}
\subfloat[AAE fraction in summary vs $\alpha$]{\hspace{.33\linewidth}}
  \subfloat[Rouge-1 F-score vs $\alpha$]{\hspace{.33\linewidth}}
\caption{Evaluation of our model on datasets containing 50\% AAE tweets using SummaRuNNer as algorithm $A$.
}
\label{fig:random_aae_frac_fair_50_srnn}
\end{figure*}

For random collections of TwitterAAE dataset, with different fraction of AAE tweets in them, we use our model to generate summaries of different sizes.
The results for TF-IDF are given in Figure~\ref{fig:random_aae_frac_fair_09_tfidf} and \ref{fig:random_aae_frac_fair_50_tfidf}; for Hybrid-TF-IDF, see Figure~\ref{fig:random_aae_frac_fair_09_hybtfidf} and \ref{fig:random_aae_frac_fair_50_hybtfidf}; for LexRank, see Figure~\ref{fig:random_aae_frac_fair_09_lexrank} and \ref{fig:random_aae_frac_fair_50_lexrank}; for TextRank, see Figure~\ref{fig:random_aae_frac_fair_09_textrank} and \ref{fig:random_aae_frac_fair_50_textrank}; for SummaRuNNer, see Figure~\ref{fig:random_aae_frac_fair_09_srnn} and \ref{fig:random_aae_frac_fair_50_srnn}.
%
%
$\alpha=0.5$, unless mentioned otherwise.
  
\subsection{Evaluation of our model on keyword-specific collections of TwitterAAE} \label{sec:aae_query_all_results}

\begin{figure*}[!htbp]
\small
\includegraphics[width=0.9\linewidth]{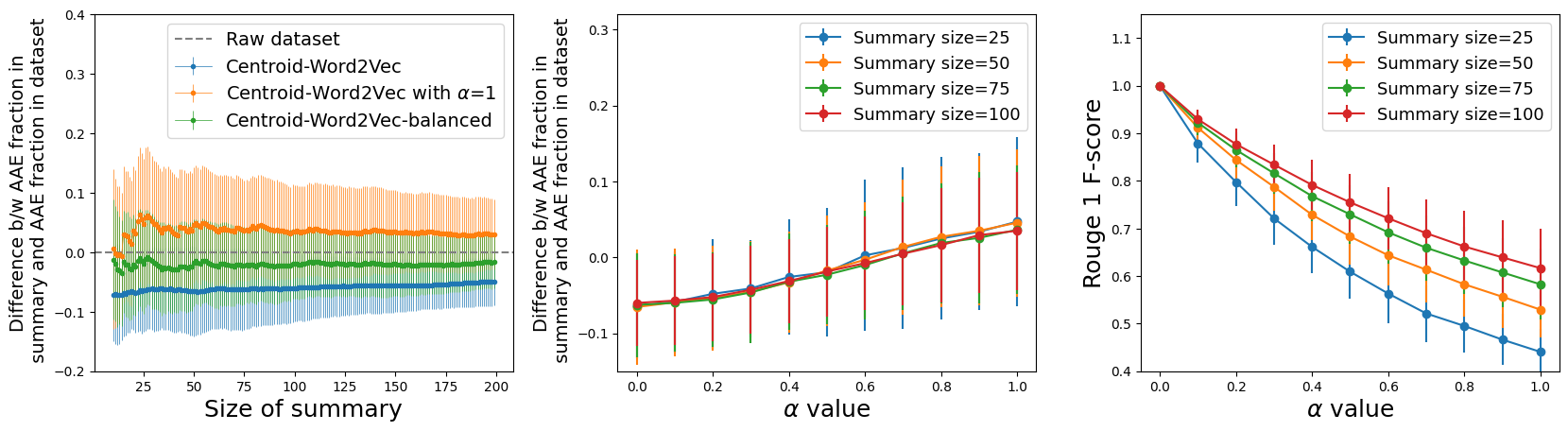}
  \subfloat[AAE fraction vs summary size]{\hspace{.33\linewidth}}
\subfloat[AAE fraction in summary vs $\alpha$]{\hspace{.33\linewidth}}
  \subfloat[Rouge-1 F-score vs $\alpha$]{\hspace{.33\linewidth}} \\
\includegraphics[width=0.9\linewidth]{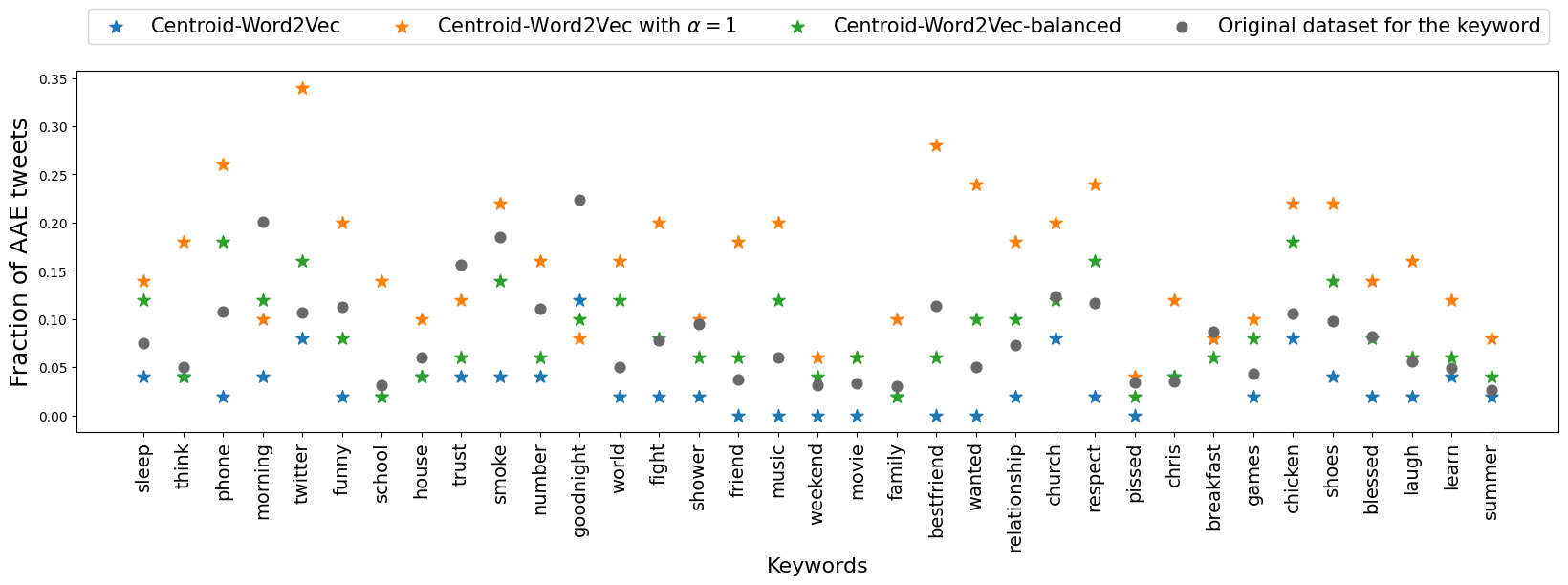}  
  \subfloat[AAE fraction for different keywords and summary size = 50]{\hspace{\linewidth}}
\caption{Evaluation of our model on keyword-specific datasets using Centroid-Word2Vec as algorithm $A$.
}
\label{fig:query_aae_frac_fair_w2v}
\end{figure*}

\begin{figure*}[!htbp]
\small
\includegraphics[width=0.9\linewidth]{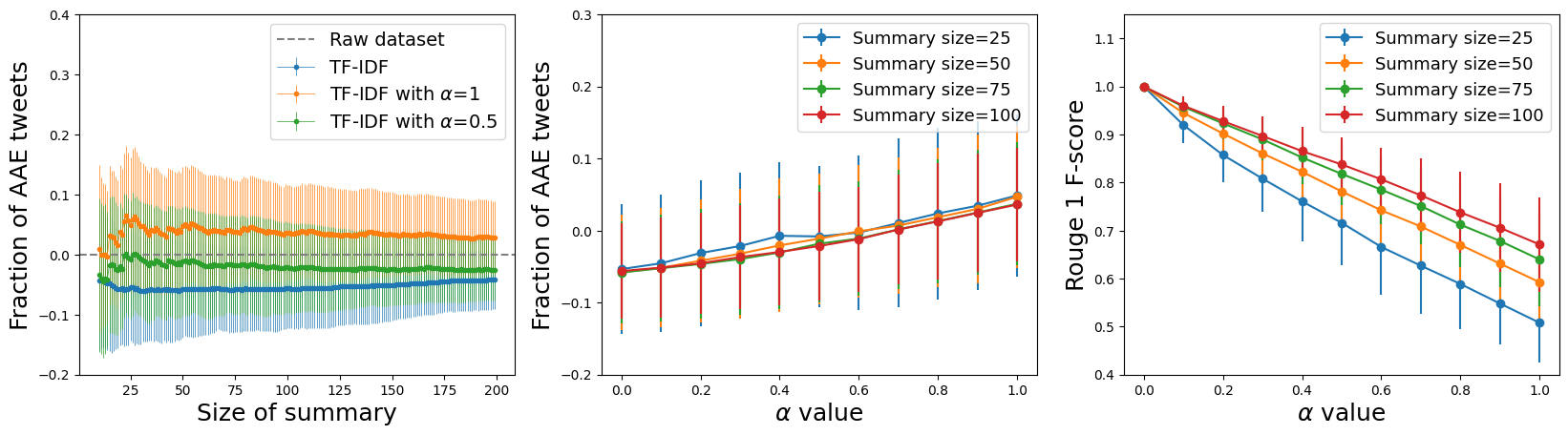}
  \subfloat[AAE fraction vs summary size]{\hspace{.33\linewidth}}
\subfloat[AAE fraction in summary vs $\alpha$]{\hspace{.33\linewidth}}
  \subfloat[Rouge-1 F-score vs $\alpha$]{\hspace{.33\linewidth}} \\
\includegraphics[width=0.9\linewidth]{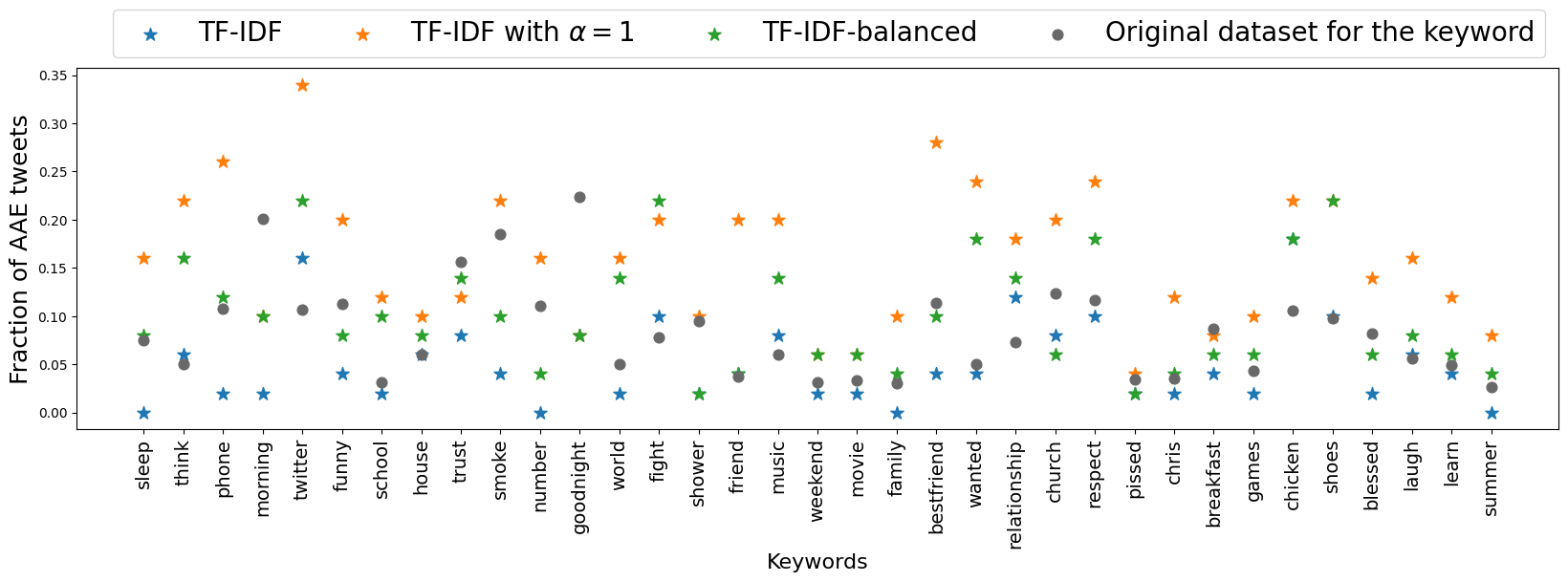}  
  \subfloat[AAE fraction for different keywords and summary size = 50]{\hspace{\linewidth}}
\caption{Evaluation of our model on keyword-specific datasets using TF-IDF as algorithm $A$.
}
\label{fig:query_aae_frac_fair_tfidf}
\end{figure*}

\begin{figure*}[!htbp]
\small
\includegraphics[width=0.9\linewidth]{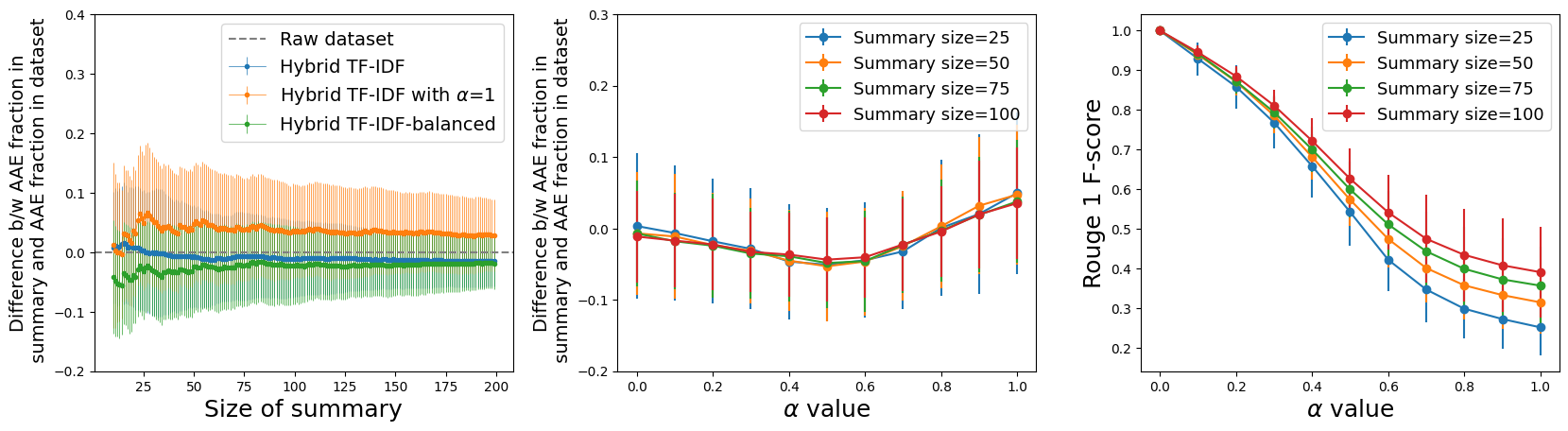}
  \subfloat[AAE fraction vs summary size]{\hspace{.33\linewidth}}
\subfloat[AAE fraction in summary vs $\alpha$]{\hspace{.33\linewidth}}
  \subfloat[Rouge-1 F-score vs $\alpha$]{\hspace{.33\linewidth}} \\
\includegraphics[width=0.9\linewidth]{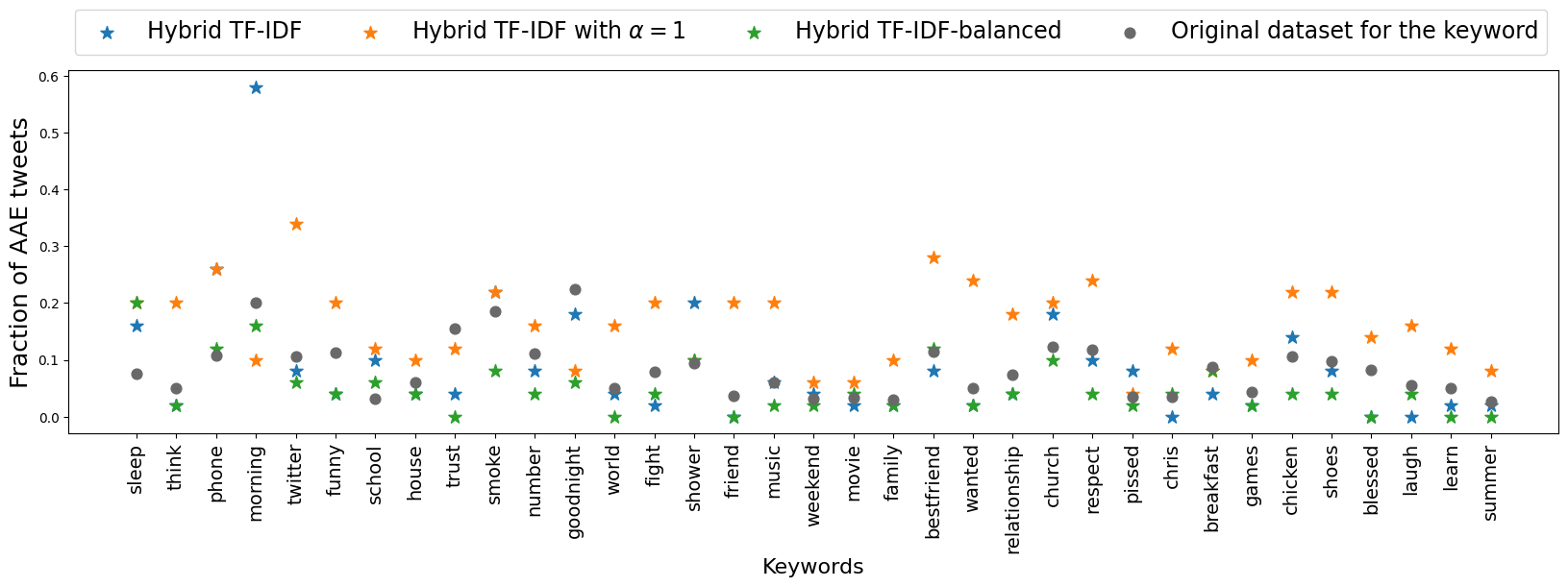}  
  \subfloat[AAE fraction for different keywords and summary size = 50]{\hspace{\linewidth}}
\caption{Evaluation of our model on keyword-specific datasets using Hybrid TF-IDF as algorithm $A$. Here $\alpha=0.7$ for balanced algorithm.
}
\label{fig:query_aae_frac_fair_hybtfidf}
\end{figure*}

\begin{figure*}[!htbp]
\small
\includegraphics[width=0.9\linewidth]{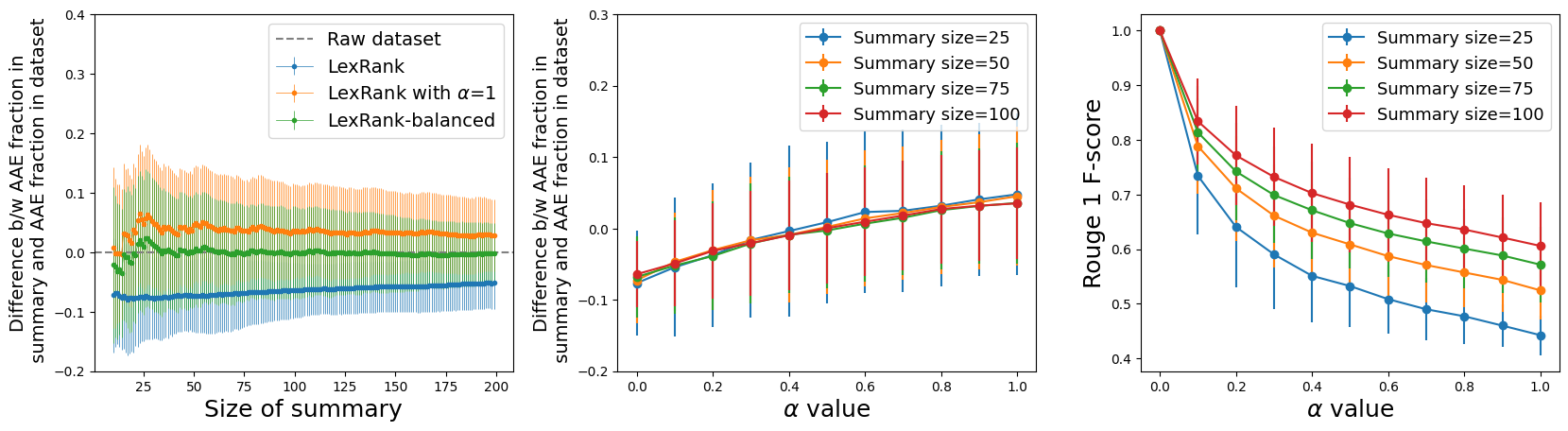}
  \subfloat[AAE fraction vs summary size]{\hspace{.33\linewidth}}
\subfloat[AAE fraction in summary vs $\alpha$]{\hspace{.33\linewidth}}
  \subfloat[Rouge-1 F-score vs $\alpha$]{\hspace{.33\linewidth}} \\
\includegraphics[width=0.9\linewidth]{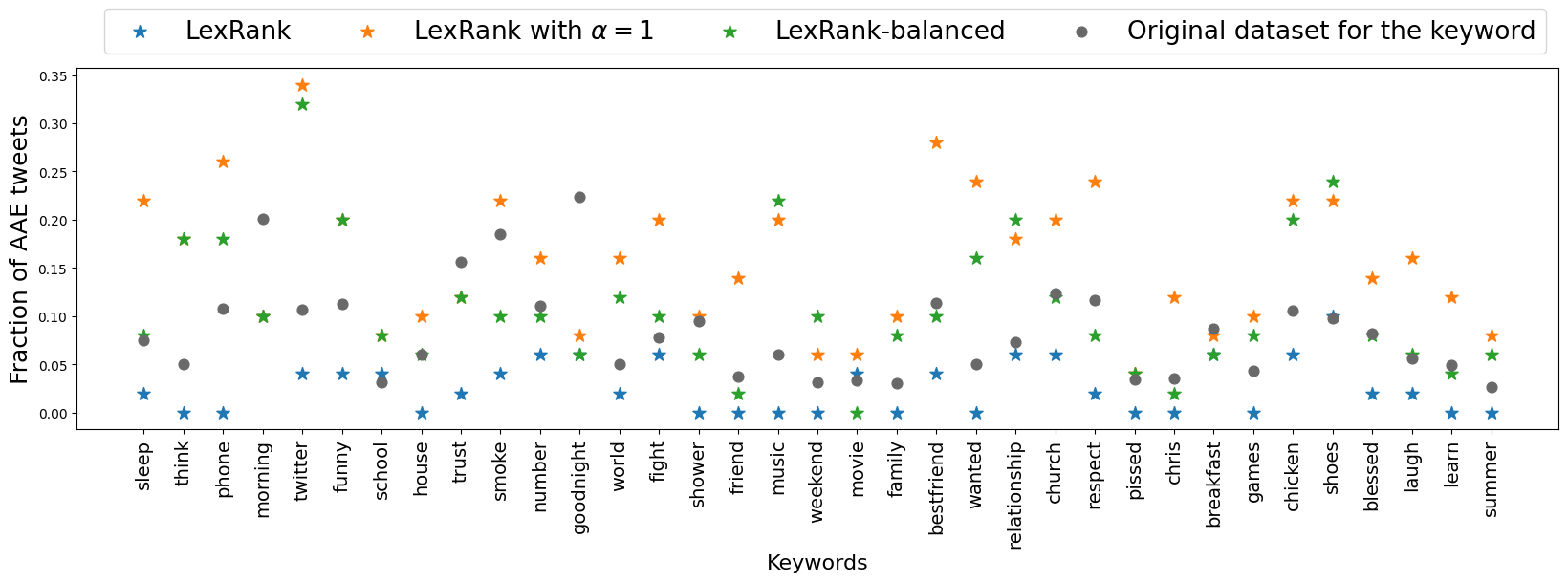}  
  \subfloat[AAE fraction for different keywords and summary size = 50]{\hspace{\linewidth}}
\caption{Evaluation of our model on keyword-specific datasets using LexRank as algorithm $A$.
}
\label{fig:query_aae_frac_fair_lexrank}
\end{figure*}

\begin{figure*}[!htbp]
\small
\includegraphics[width=0.9\linewidth]{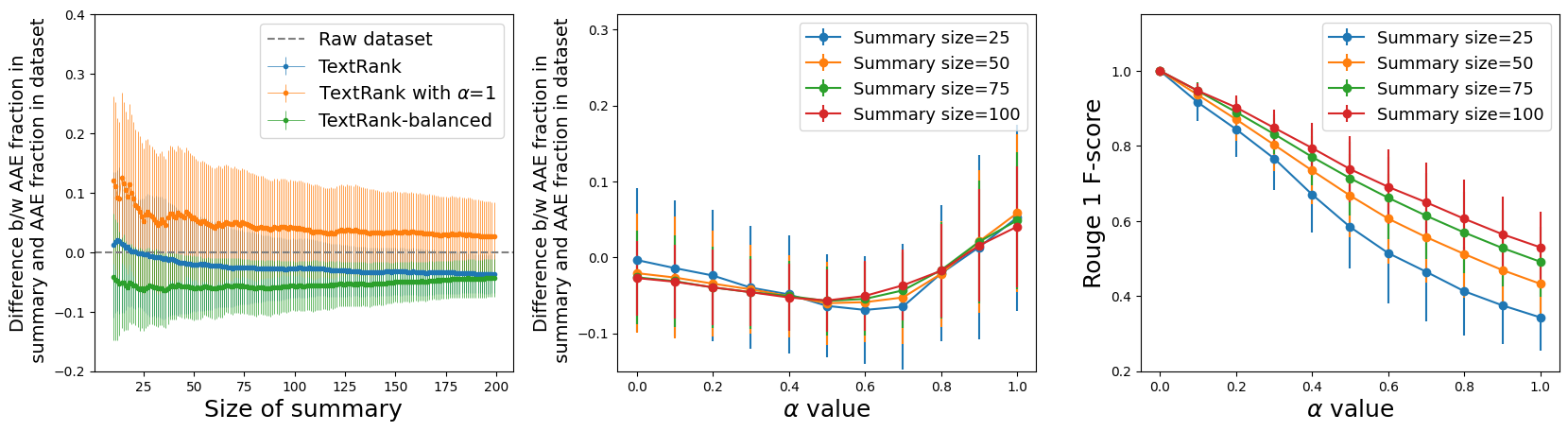}
  \subfloat[AAE fraction vs summary size]{\hspace{.33\linewidth}}
\subfloat[AAE fraction in summary vs $\alpha$]{\hspace{.33\linewidth}}
  \subfloat[Rouge-1 F-score vs $\alpha$]{\hspace{.33\linewidth}} \\
\includegraphics[width=0.9\linewidth]{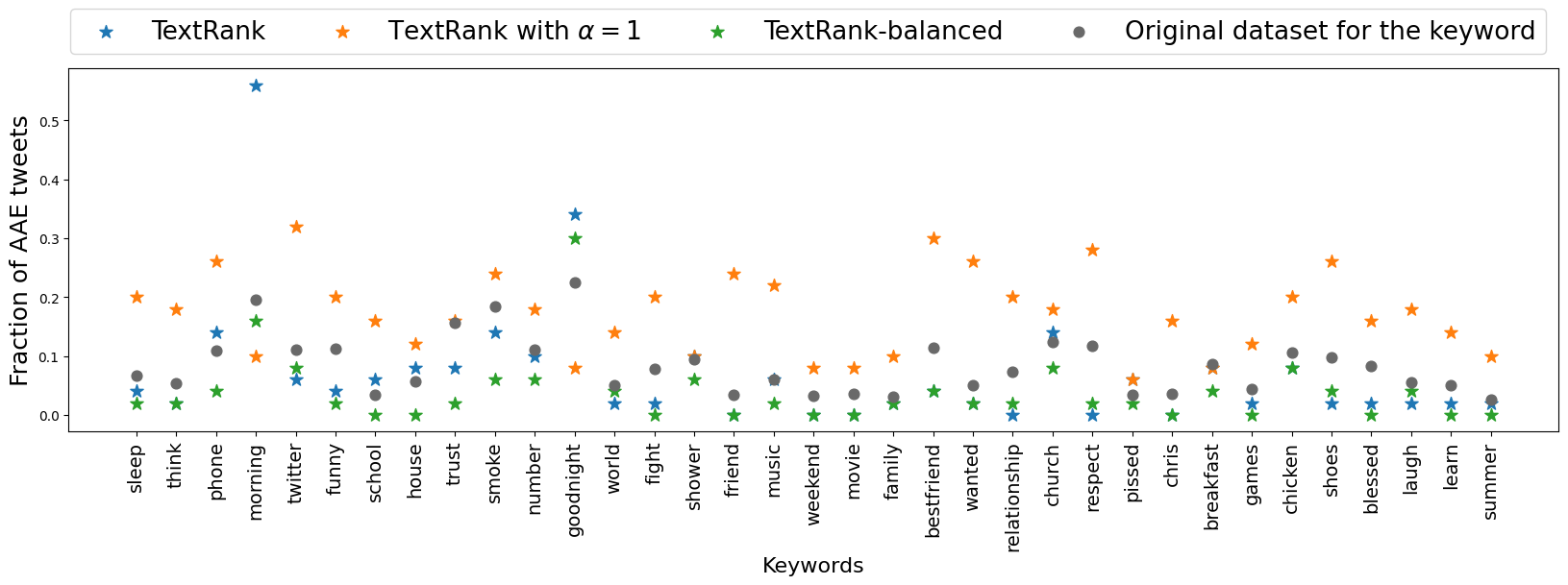}  
  \subfloat[AAE fraction for different keywords and summary size = 50]{\hspace{\linewidth}}
\caption{Evaluation of our model on keyword-specific datasets using TextRank as algorithm $A$. Here $\alpha=0.7$ for balanced algorithm.
}
\label{fig:query_aae_frac_fair_textrank}
\end{figure*}

\begin{figure*}[!htbp]
\small
\includegraphics[width=0.9\linewidth]{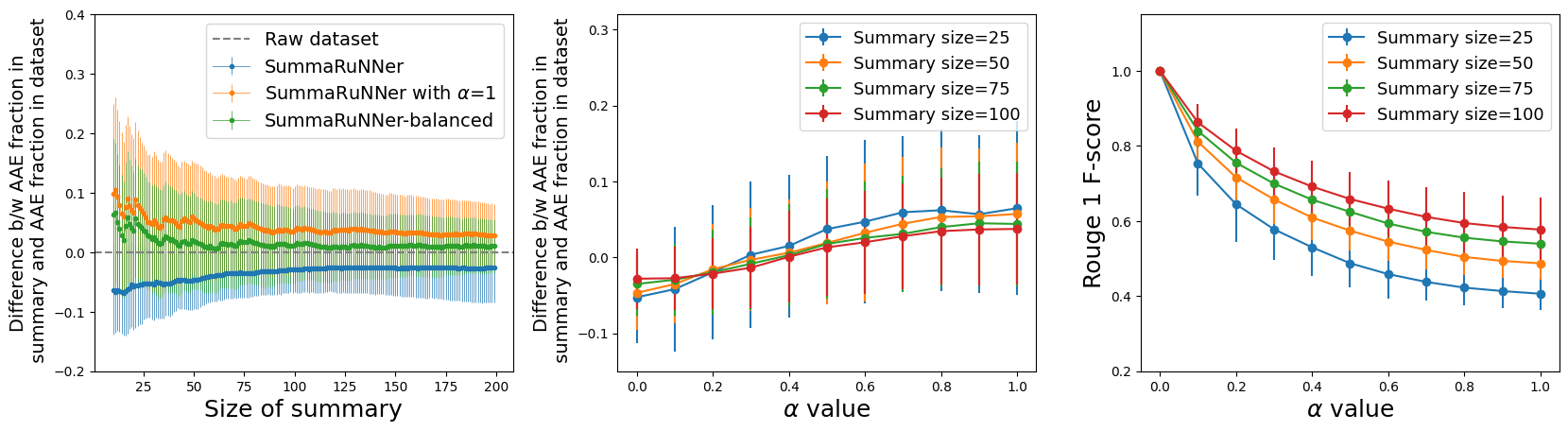}
  \subfloat[AAE fraction vs summary size]{\hspace{.33\linewidth}}
\subfloat[AAE fraction in summary vs $\alpha$]{\hspace{.33\linewidth}}
  \subfloat[Rouge-1 F-score vs $\alpha$]{\hspace{.33\linewidth}} \\
\includegraphics[width=0.9\linewidth]{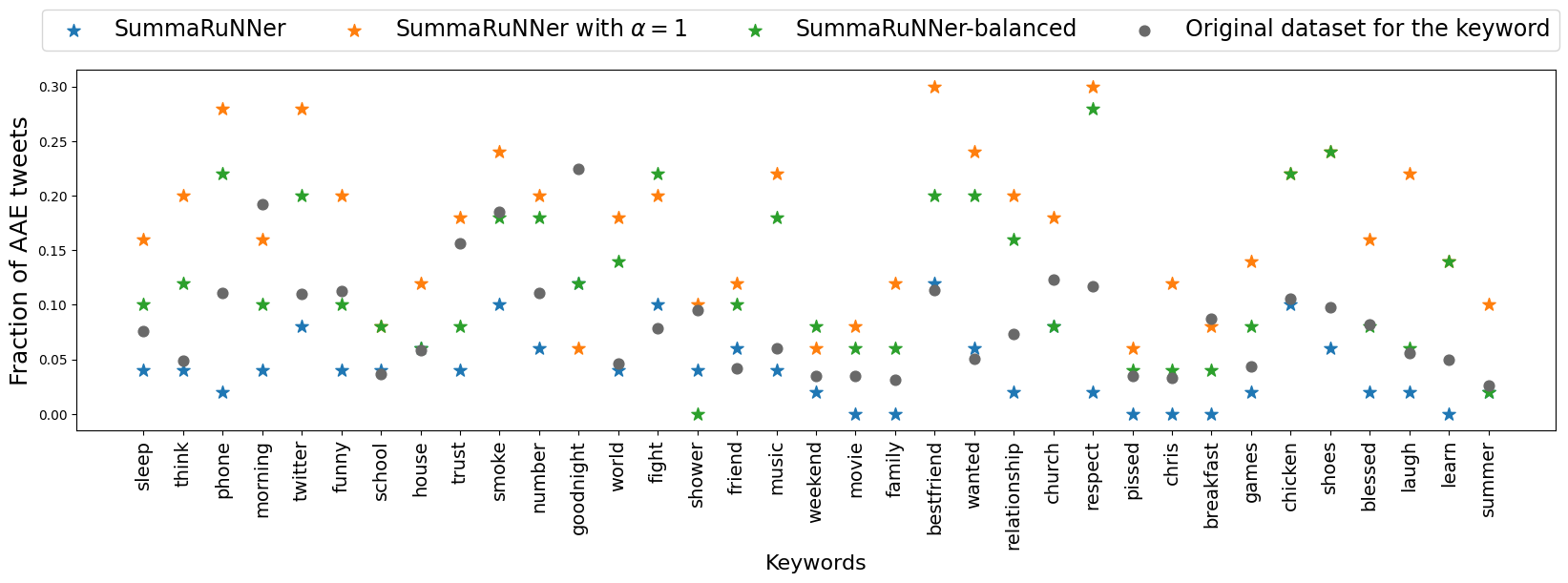}  
  \subfloat[AAE fraction for different keywords and summary size = 50]{\hspace{\linewidth}}
\caption{Evaluation of our model on keyword-specific datasets using SummaRuNNer as algorithm $A$.
}
\label{fig:query_aae_frac_fair_srnn}
\end{figure*}

Next, we also present the results for our model on collections of TwitterAAE dataset containing the keywords used in Section~\ref{sec:orig_expts}.
The results for TF-IDF are given in Figure~\ref{fig:query_aae_frac_fair_tfidf}; for Hybrid-TF-IDF, see Figure~\ref{fig:query_aae_frac_fair_hybtfidf}; for LexRank, see Figure~\ref{fig:query_aae_frac_fair_lexrank}; for TextRank, see Figure~\ref{fig:query_aae_frac_fair_textrank}; for Centroid-Word2Vec, see Figure~\ref{fig:query_aae_frac_fair_w2v}; for SummaRuNNer, see Figure~\ref{fig:query_aae_frac_fair_srnn}.

\subsection{Evaluation of our model using different diversity set compositions}
\begin{figure*}[h]
\small
\includegraphics[width=0.9\linewidth]{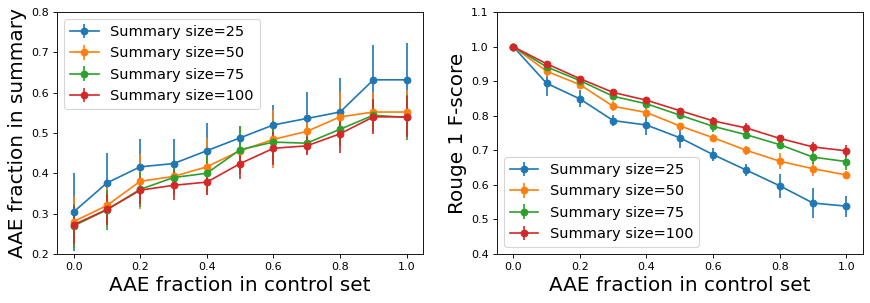}
  \subfloat[AAE fraction in summary vs control set]{\hspace{.5\linewidth}}
  \subfloat[Rouge-1 F-score vs control set]{\hspace{.5\linewidth}} \\
\caption{Evaluation of our model using different control set compositions.
}
\label{fig:random_aae_frac_fair_w2v_diff_control}
\end{figure*}

We also present the evaluation for the setting where the diversity control set has unequal fraction of AAE and WHE posts.
For random collections where the fraction of AAE posts in collection is 50\%, Figure~\ref{fig:random_aae_frac_fair_w2v_diff_control}. 
As expected, the fraction of AAE posts in summary increases as fraction of AAE posts in control set increases.
This is another parameter that can be tuned to adjust and obtain the desired fraction of AAE posts in the summary.

\section{Other details and results for Crowdflower Gender AI dataset} \label{sec:gender_appendix}

The diversity control set used for Crowdflower Gender evaluation is presented in Table~\ref{tbl:div_control_gender}.

\begin{table}[!htbp]
  \caption{Diversity control set for simulations on Crowdflower Gender AI dataset} 
  \small
\begin{tabular}{|p{\linewidth}|}
    \hline\\
     \textbf{Tweets by female user-accounts}  \\
     \hline \\
``jameslykins haha man! the struggle is reeeeeal! '' \\
``red lips and rosy cheeks'' \\
``\#mood spirit of jezebel control revelation 21820, 26 a war goes on in todays church, and the '' \\
``where the hell did october go? halloween is already this weekend. '' \\
``my lipstick looked like shit and my hair is usually a mess but im still cute tho so  '' \\
``say she gon ride for me , ill buy the tires for you  '' \\
``so excited to start the islam section in my religions class    '' \\
``wow blessed my 200 kate spade bag is ripping and ive only used it twice a week since the end of september .'' \\
``all ive done today is lie around and homework tbh'' \\
``of course you want to blame me for not finishing college and thus bringing this debt to myself of course'' \\
``misskchrista everyone was obsessed with rhys though, no one really knew the other two xxx'' \\
``papisaysyes at first i thought this said, my dick is on drugs and i still dont know which is worse lol'' \\
``huge announcement and \#career change for 2016. \#goals \#dreams \#nymakeupartist '' \\
``practice random acts of kindness and make it a habit \#aldubpredictions'' \\
`` sammanthae glad i can make you laugh i miss you and love you too!!'' \\
``nba i play basketball to escape reality. between the exercise and the diff personalities memories are made!'' \\
``z100newyork please let me attend the future now vip party tonight i love demi and nick \#z100futurenow  '' \\
``\#win 2 random jumbies stuffed animals \#giveaway us only 1113 bassgiraffe '' \\
``daynachirps thats a great point. thanks for the reminder. \#contentchat'' \\
``ive told bri all this time it would happen and it finally did '' \\
    \hline 
    \hline \\
     \textbf{Tweets by male user accounts}  \\
     \hline \\
``warrenm ill be using my new mbp. i do see dells 5k line needs 2 thunderbolt connections to make it a true 5k display. not the case here?'' \\
``logic301 salute on the new visuals my g! dope as fuck'' \\
``i liked a youtube video official somewhere over the rainbow 2011 israel iz kamakawiwoole'' \\
``laughs and cries at the same time cause true '' \\
``akeboshi night and day'' \\
``now you all know the monster mash, but now for something really scary, the climate mash '' \\
``i hate when u tell someone u love them and they ignore u  '' \\
``the finger hahsah '' \\
``the corruption of the wash. d.c. crowd is now of epic proportions. enlist gt join us '' \\
``i wish i went to school closer to mark a schwab . beating up doors and walls looks like a lot of fun.'' \\
``keepherwarm kobrakiddlng aimhbread now ill let you know that ive known a guy my whole life who dated several girls and then later on'' \\
``xavierleon  fr like wtf are they taking that they just cant fucking dye and busting through doors?!  '' \\
``heh, i just remember people actually think that se and hp are intentionally sabotaging the football team.'' \\
``we must lessen the auditory deprivation! i agree earlier the implantation, the better! '' \\
``\#repost seekthetruth with repostapp. repost  ugly by nature 85 of the \#tampons, cotton and '' \\
``the \#ceo needs to embrace and sell social to the team or else is goes nowhere. bernieborges \#h2hchat \#ibminsight '' \\
``if you scored a touchdown on sunday and didnt dab, hit them folks, or do that hotline bling dance, it shouldnt have counted.'' \\
``zbierband yo zbb, played our last seasonal gig at st. jude. good times had by all. remember the more you drink, the better we sound!'' \\
``i hate writing on the first page of a notebook i feel like im ruining something so perfect'' \\
``we schools should be given credit for growth in the apr, but growth is not the destination. michael jones moboe. '' \\
  \hline
  \end{tabular}
  \label{tbl:div_control_gender}
\end{table}

\subsection{Evaluation of our model with different blackbox algorithms}
\begin{figure*}[!htbp]
\small
\includegraphics[width=\linewidth]{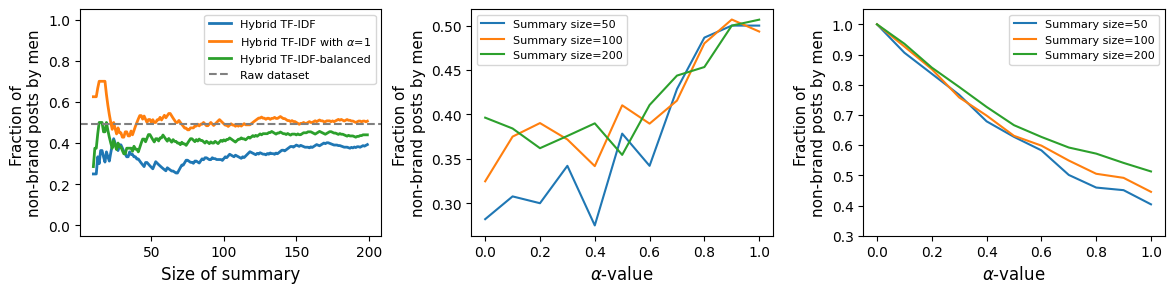}
  \subfloat[Gender fraction vs summary size]{\hspace{.33\linewidth}}
\subfloat[Gender fraction vs $\alpha$]{\hspace{.33\linewidth}}
  \subfloat[Rouge-1 F-score vs $\alpha$]{\hspace{.33\linewidth}}
\caption{Evaluation of our model on Crowdflower Gender AI dataset using Hybrid TF-IDF as algorithm $A$.
}
\label{fig:gender_fair_hybtfidf}
\end{figure*}

\begin{figure*}[!htbp]
\small
\includegraphics[width=\linewidth]{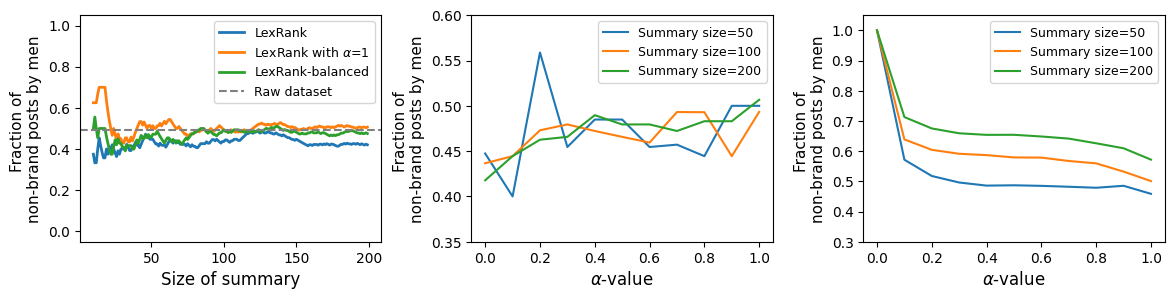}
  \subfloat[Gender fraction vs summary size]{\hspace{.33\linewidth}}
\subfloat[Gender fraction vs $\alpha$]{\hspace{.33\linewidth}}
  \subfloat[Rouge-1 F-score vs $\alpha$]{\hspace{.33\linewidth}}
\caption{Evaluation of our model on Crowdflower Gender AI dataset using LexRank as algorithm $A$.
}
\label{fig:gender_fair_lexrank}
\end{figure*}

\begin{figure*}[!htbp]
\small
\includegraphics[width=\linewidth]{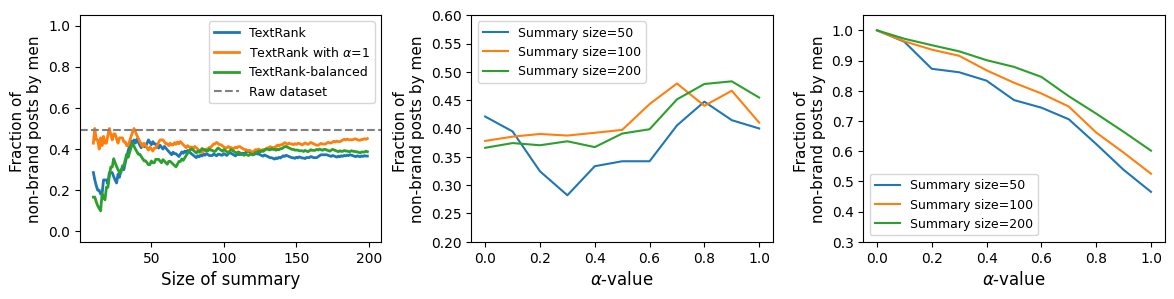}
  \subfloat[Gender fraction vs summary size]{\hspace{.33\linewidth}}
\subfloat[Gender fraction vs $\alpha$]{\hspace{.33\linewidth}}
  \subfloat[Rouge-1 F-score vs $\alpha$]{\hspace{.33\linewidth}}
\caption{Evaluation of our model on Crowdflower Gender AI dataset using TextRank as algorithm $A$.
}
\label{fig:gender_fair_textrank}
\end{figure*}

\begin{figure*}[!htbp]
\small
\includegraphics[width=\linewidth]{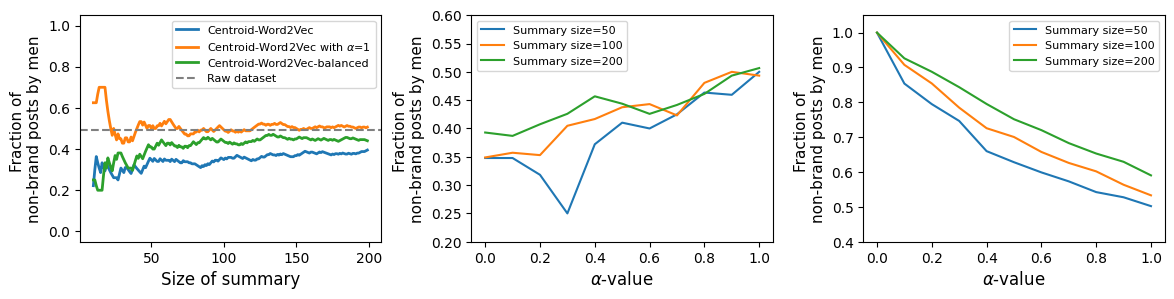}
  \subfloat[Gender fraction vs summary size]{\hspace{.33\linewidth}}
\subfloat[Gender fraction vs $\alpha$]{\hspace{.33\linewidth}}
  \subfloat[Rouge-1 F-score vs $\alpha$]{\hspace{.33\linewidth}}
\caption{Evaluation of our model on Crowdflower Gender AI dataset using Centroid-Word2Vec as algorithm $A$.
}
\label{fig:gender_fair_w2v}
\end{figure*}

The performance of our model using different blackbox algorithms is presented here.
The results for Hybrid TF-IDF are given in Figure~\ref{fig:gender_fair_hybtfidf};  for LexRank, see Figure~\ref{fig:gender_fair_lexrank}; for TextRank, see Figure~\ref{fig:gender_fair_textrank}; for Centroid-Word2Vec, see Figure~\ref{fig:gender_fair_w2v}.

\newpage

\section{Diversity control set used for Claritin dataset}

\begin{table}[!htbp]
  \caption{Diversity control set for simulations on Claritin dataset} 

  \small
\begin{tabular}{|p{\linewidth}|}
    \hline\\
     \textbf{Tweets by female user-accounts}  \\
     \hline \\
``claritin, why didnt you work? i was desperate thats why i took you. ang mahal mo pa man din! ''\\
 ``ATMENTION been there. always. done that. youll be fine. claritin works for that.   ''\\
 ``ATMENTION all allergy meds raise als blood pressure a lot. claritin isnt so bad but still sucks. the kid stuff is half dose and works ''\\
 ``k time to bust out that claritin. siiigh ''\\
 ``ATMENTION if they are asking for allegra, mucinex, or claritin. they want the d.   ATMENTION ''\\
 ``if a girl sends you a text, heyy, im sick. . she probably wants the d  claritin d \#pervs ''\\
 ``ATMENTION yes, claritin, tylenol and ibuprofen. ''\\
 ``what ever happened to jeff corwin? supposedly he does claritin commercials now. ''\\
 ``deffo allergic to tingle creams now not on my legs back or belly though but on my arms chest ampface need to buy claritin amp chamomile lotion ''\\
 ``ATMENTION awesome i never wear glasses so this has sucked\! doc said taking one claritin dried up my tears. just one?? ''\\    \hline 
    \hline \\
     \textbf{Tweets by male user accounts}  \\
     \hline \\
 ``ATMENTION i have one xd  if she has allergies.. give here some claritind ! ''\\
  ``if a girl tells you shes sick she wants the d, claritind ATMENTION ''\\
  ``ATMENTION givin complementary claritin d pills amp shit. ''\\
  ``claritin and food please \#sniffle ''\\
  ``ok so 2 pills of allegra is not helping my allergies, anyone have another pill i should try? claritin is out ''\\
  ``she feeling sick? she wants the d. claritind ''\\
  ``yeah my allergies are acting up , i didnt take any claritin today  ATMENTION ''\\
  ``ATMENTION if a girl sends you hey, im sick. she probably wants the claritind. haha.      ''\\
  ``clearly. claritin clear ''\\
  ``her allergies were acting up, so i gave her the d.... claritin d. ''\\  \hline
  \end{tabular}
  \label{tbl:div_control_claritin}
\end{table}

\end{document}